\documentclass[aps,prb,reprint,showpacs,floatfix]{revtex4-1}
\usepackage{amsmath,amssymb,amsfonts,graphicx,placeins,float,bm}
\usepackage{hhline,enumerate}
\usepackage{hyperref}
\usepackage{blindtext}
\usepackage{xcolor}
\newcommand{\bs}{\boldsymbol}

\graphicspath{{Figs/}}

\begin{document}

\title{Spectral properties of superconductors with ferromagnetically ordered magnetic impurities}

\author{Daniel Persson}
\email{daniel.j.persson@chalmers.se}
\author{Oleksii Shevtsov}
\email{shevtsov@chalmers.se}
\author{Tomas L\"{o}fwander}
\email{tomas.lofwander@chalmers.se}
\author{Mikael Fogelstr\"{o}m}
\email{mikael.fogelstrom@chalmers.se}

\affiliation{Department of Microtechnology and Nanoscience -- MC2, Chalmers University of Technology, SE-412 96 G\"{o}teborg, Sweden}

\date{\today}

\pacs{74.78.-w, 74.78.Na, 74.20.Fg, 74.62.Dh, 73.63.-b}

\begin{abstract}

We present a comprehensive theoretical study of thermodynamic properties of superconductors with a dilute concentration of magnetic impurities,
with focus on how the properties of the superconducting host change if the magnetic moments of the impurities order ferromagnetically.
Scattering off the magnetic impurities leads to the formation of a band of Yu-Shiba-Rusinov states within the superconducting energy gap
that drastically influences superconductivity.
In the magnetically ordered system, the magnetization displays a sudden drop as function of impurity density or magnetic moment amplitude.
The drop occurs as the spin-polarized impurity band crosses the Fermi level and is associated with
a quantum phase transition first put forward by Sakurai for the single impurity case.
Taking into account that the background magnetic field created by the ordered impurity moments enters as a Zeeman shift,
we find that the superconducting phase transition changes from second order to first order for high enough impurity concentration.

%We consider a superconductor with a dilute concentration of magnetic impurities. The impurities are described by the Yu-Shiba-Rusinov model. We investigate how properties of the superconducting host change if we allow the magnetic moments of impurities to be ferromagnetically ordered. We show that this case has many new features compared to the case of unpolarized impurities. In particular, it is necessary to take into account a background magnetic field created by the impurity moments in order to get physically meaningful results. We find that the superonducting phase transition changes from the second-order to the first-order for high enough impurity concentrations and \textcolor{red}{it is not enough to solve the gap equation to find the correct transition temperature. This case is analyzed by computing the free energy difference in superconducting and normal states.} We also compute a magnetization in the system as a function of impurity density and magnetic moment amplitude, and demonstrate \textcolor{red}{a signature of} the quantum phase transition put forward by A.~Sakurai. Finally we compute a differential Fano factor for the tunneling current, which enables us to elucidate the relative role of single-particle and Andreev scatterings in the tunneling spectra of opposite-spin channels.
\end{abstract} 

\maketitle

\section{Introduction}
In conventional and most known high-temperature superconductors Cooper pairs are formed of electrons with anti-parallel spins (spin-singlets). On the other hand, existence of a ferromagnetic order favors parallel spin alignment. Therefore superconductivity and ferromagnetism are two competing orders and they are known to coexist in only a few bulk rare-earth materials \cite{Felner1997,Saxena2000,Aoki2001,Pfleiderer2001} and more recently at an interface between two bulk insulators, $\mathrm{LaAlO_3}$ (LAO) and $\mathrm{SrTiO_3}$ (STO) \cite{Dikin2011}. Though such systems are rare in Nature, they can be engineered artificially in superconducting heterostructures. An example of such systems are ferromagnet-superconductor (FS) interfaces and SFS Josephson junctions, where a long-range proximity effect was observed \cite{Keizer2006,Anwar2010,Robinson2010}. Another type of system which are a focus of active research are superconductors with magnetic impurities. The term ``magnetic impurity" in this context may refer to a single transition or rear-earth atom \cite{Woolf1965,Yazdani1997,Hudson2001,Ji2008} or small ferromagnetic islands \cite{Asulin2009}. Another reason for the current interest in these systems is the search for experimental evidence of elusive Majorana particles. According to theoretical predictions \cite{Nadj-Perge2013,Pientka2013,Pientka2013,Peng2015} they emerge at the ends of linear chains of magnetic impurities, and are believed to have been observed in recent experiments \cite{Nadj-Perge2014}.

From a theoretical point of view, investigation of properties of superconductors with magnetic impurities dates back to the work by Abrikosov and Gorkov \cite{Abrikosov_Gorkov_1961}. Within the first order Born approximation (weak impurity scattering) they demonstrated that magnetic impurities lead to pair-breaking and within certain parameter regimes of the model, gapless superconductivity can emerge. Taking into account the scattering off a single impurity exactly, to all orders in perturbation theory, later it was shown \cite{Yu1965,Shiba1968,Rusinov1969} that magnetic impurities in a superconductor are able to host single-particle bound states \cite{Zittartz1972,Tsang1980,Bauriedl1981,Balatsky2006,Kim2015}, commonly known now as Yu-Shiba-Rusinov (YSR) states. Each YSR state is spin-polarized and does not have a Kramers partner due to explicitly broken time-reversal symmetry. As a consequence, occupying or emptying such a state, the system undergoes a quantum phase transition and its ground state changes its parity, at the same time gaining or loosing one single-particle spin. Since this idea was first put forward by Sakurai \cite{Sakurai1970}, to our knowledge it received a very limited attention resulting in a few published theoretical works \cite{Salkola1997,Morr2003,Morr2006}. In these studies they usually performed self-consistent tight-binding calculations with one, two or three impurities to demonstrate the phase transition and spin-polarization of the ground state. We have to note that YSR states as discussed above were found by treating magnetic impurity spins as classical. They emerge in a superconductor as an attempt to screen the impurity magnetic moment, similarly to the Kondo effect in normal metals \cite{Kondo1964}. On the other hand, treating impurity spins quantum mechanically, it is possible to study the interplay of Kondo screening and superconductivity \cite{Liu1965,Griffin1965,Maki1967,Soda1967,Fowler1967,Abrikosov1969,Takano1969,Kitamura1970,Fowler1970,MullerHartmann1971,Matsuura1977,Ichinose1977,Ichinose1977_2,Franke2011}, however theoretical treatment of such systems becomes much more involved.

Consider a superconductor with a finite concentration of YSR magnetic impurities. An open question is how superconducting properties are altered when the impurity spins become ferromagnetically ordered. This can happen, for example, due to an external magnetic field, impurity exchange interaction (direct or mediated by the itinerant electrons) or an intrinsic magnetic anisotropy. This issue was studied theoretically \cite{Gorkov_Rusinov_1964,Fulde_Maki_1966,Izyumov1974} treating impurity scattering in the Born limit. In this paper we extend this study by describing the single-impurity scattering exactly, using the t-matrix formulation \cite{Yu1965,Shiba1968,Rusinov1969,Rammer_book1998}. It is important to note that previous studies based on the first order Born limit took into account only local exchange scattering of itinerant electrons off magnetic impurities. We argue that, when generalizing the model to the t-matrix approximation, it is necessary to also take into account a background magnetic field created by the impurity spins in order to obtain physically sensible results. This circumstance forbids considering the unitary limit of impurity scattering, since increasing the magnitude of magnetic moments (unitary limit corresponds to letting the moments be infinitely large) drives the system to the normal state quickly. In the framework of the qusiclassical Green's functions formalism \cite{Eilenberger1968,Larkin_Ovchinnikov_1969,Eschrig_PRB2000,Eschrig_PRB2009} we study the behavior of the order parameter and the superconducting transition temperature for the case of unpolarized and ferromagnetically ordered impurities. For the ordered case we find that for high impurity concentrations the order of phase transition can change from second to first.
%On the other hand, the self-consistency equation for the order parameter can have more than one solution and, in order to single out the physically relevant one, we compute a difference between Gibbs free energies in superconducting and normal states. When the phase transition is first-order superconducting transition temperature occurs when the free energy difference crosses zero. All these features are absent in the case of unpolarized impurities.
Furthermore, by computing the magnetization in the system (in the ordered case) we are able to demonstrate the signature of the quantum phase transition put forward by Sakurai \cite{Sakurai1970} for a finite concentration of impurities. 

The paper is organized as follows. In Sec. \ref{Sec_Model} we describe in detail the theoretical model of the system and briefly introduce the quasiclassical Green's function formalism. We provide a set of self-consistency equations and expressions for thermodynamic quantities and observables for the two cases: (i) unpolarized and (ii) ferromagnetically ordered impurities. In Sec. \ref{Sec_Res} we use the results of the previous section to compute self-consistently the transition temperature, order parameter, density of states, and magnetization. Section \ref{Sec_Concl} contains a discussion about the validity of our model and connection to its potential experimental realization. This section also contains a summary of our findings and concludes the paper. A few technical details of the calculation have been collected in the Appendix.

\section{Theoretical Model}\label{Sec_Model}

\subsection{Description of the setup}

\begin{figure}[t]
\includegraphics[width=\columnwidth]{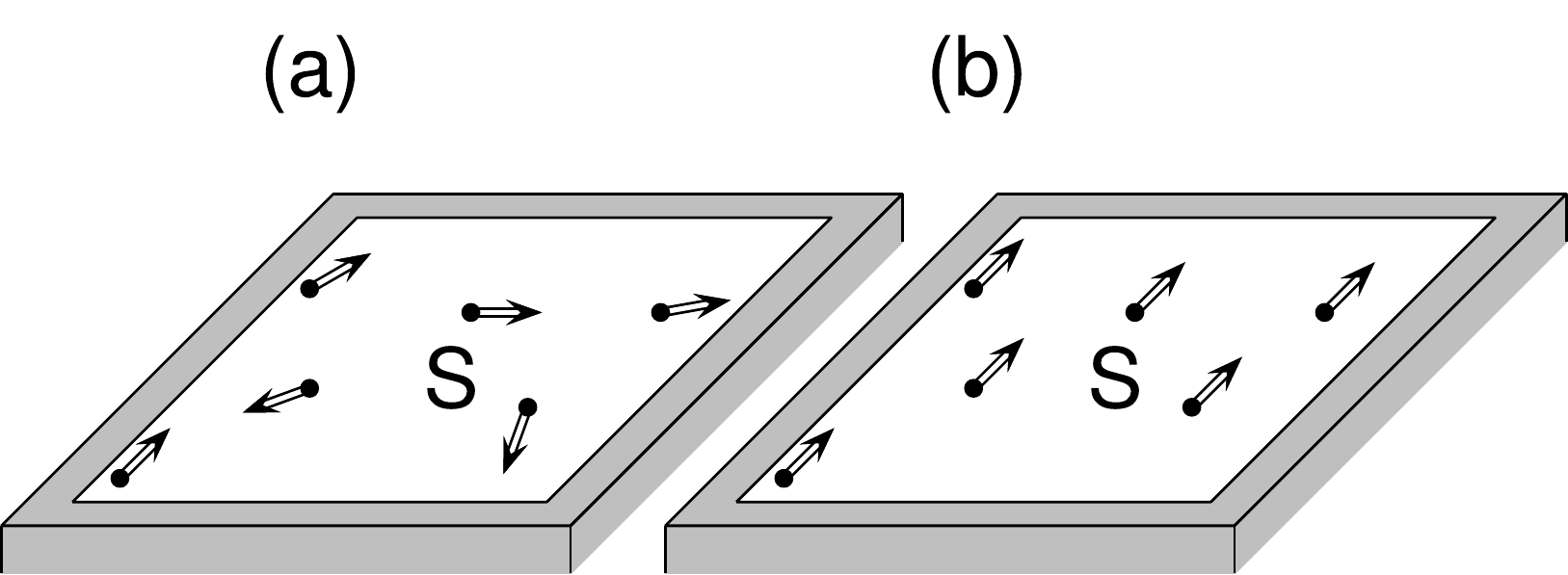}
\caption{We consider a superconducting thin film (S) deposited on a substrate. Magnetic impurities are homogeneously distributed within the sample. Magnetic moments of the impurities are either completely unpolarized (a), or ferromagnetically ordered (b) in the plane of the film.} \label{Fig1}
\end{figure}

\begin{center}
\begin{figure}[t]
\includegraphics[scale=0.5]{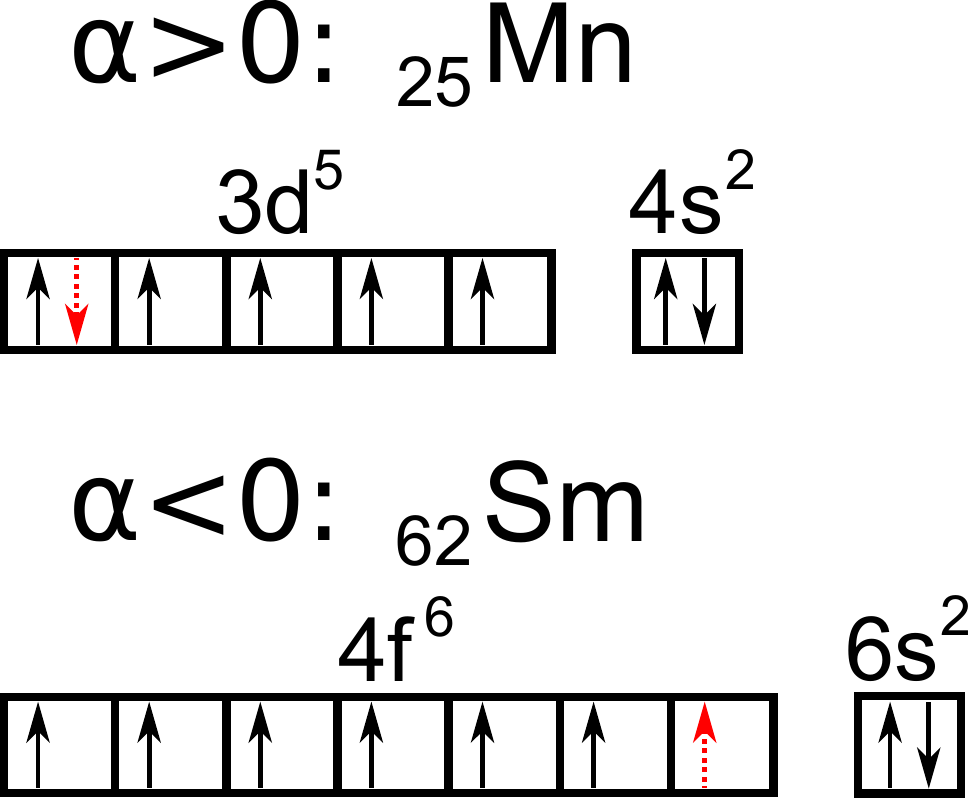}
\caption{(Color online) Example of a transition metal element (Mn) and a rear earth element (Sm) with their valence electron configurations. When placed in a superconductor, these elements strip the outer $4s^2$ and $6s^2$ shells and behave as localized magnetic moments. The Hund's rule determines the type of local exchange interaction (ferromagnetic or anti-ferromagnetic), which enters the theory through the sign of the parameter $\alpha$. Itinerant electron is depicted by the dotted red arrow.} \label{Fig2}
\end{figure}
\end{center}

The system we have in mind consists of an $s$-wave spin-singlet superconducting film with randomly distributed magnetic impurities, see Fig. (\ref{Fig1}).
The magnetic impurities are treated as classical spins, similarly to the model proposed by Yu, Shiba and Rusinov \cite{Yu1965,Shiba1968,Rusinov1969}. We consider two cases: impurity spins are (i) randomly oriented or (ii) ferromagnetically ordered in the plane of the superconducting film.
The film thickness is smaller than the London penetration depth, so that the magnetic field created in the ferromagnetically ordered case
does not lead to screening currents. The Zeeman shifts can however be substantial and we take these into account.
We describe the impurities by the following Hamiltonians
\begin{align}
&\mathcal{H}^{\mathrm{rand}}_{\mathrm{imp}} = \sum\limits_{j=1}^{N}(v_0 + \alpha v_{\mathrm{S}}\mathbf{m}_j\cdot\bs{\sigma})\delta(\mathbf{r}-\mathbf{r}_j),\label{Himp_rand}\\
&\mathcal{H}^{\mathrm{ferro}}_{\mathrm{imp}} = \beta n v_{\mathrm{S}}\mathbf{m}\cdot\bs{\sigma} + \sum\limits_{j=1}^{N}(v_0 + \alpha v_{\mathrm{S}}\mathbf{m}\cdot\bs{\sigma})\delta(\mathbf{r}-\mathbf{r}_j), \label{Himp_ferro}
\end{align}
corresponding to the two cases mentioned above. Local scattering off a given impurity at position $\mathbf{r}_j$ consists of a scalar part parametrized by $v_0$ and an exchange part. The local exchange scattering is parametrized by a dimensionless parameter $\alpha$, the tunneling amplitude of a quasiparticle onto the impurity site, and $v_{\mathrm{S}}$ the parameter proportional to the impurity magnetic moment \footnote{Because of the point-like scattering potential of impurities $v_{\mathrm{S}} = 2\tilde{v}_{\mathrm{S}}/3$, where $\tilde{v}_{\mathrm{S}} = g\mu_B\mu_0\mathcal{M}/2$. Here $g$ is the quasiparticle g-factor, $\mu_B$ is the Bohr magneton, $\mu_0$ is the vacuum permeability, and $\mathcal{M}$ is the magnitude of the impurity magnetic moment. The factor $2/3$ comes from taking into account the $\mathbf{H}$-field of a point-like magnetic dipole.}. The unit vector $\mathbf{m}_j$ points in the direction of the impurity magnetic moment, and $\bs{\sigma} = (\sigma_x,\sigma_y,\sigma_z)^T$ is the vector of Pauli matrices in spin space. For the case of unpolarized magnetic impurities, we have $\sum_{j=1}^{N}\mathbf{m}_j = 0$. The parameter $|\alpha| \leq 1$ can take both positive and negative values depending on the microscopic nature of impurities, see Fig. \ref{Fig2}. For the case of impurities made of transition metal atoms with partially filled electronic d-shells, e.g. manganese Mn, the local exchange interaction with itinerant electrons in the superconductor is anti-ferromagnetic, $\alpha > 0$ \footnote{It might seem counterintuitive that anti-ferromagnetic interaction is described by $\alpha > 0$, however it is easily understood. Magnetic moment of an electron is $\bs{\mu}_e = -g\mu_B\bs{\sigma}/2$, while its spin angular momentum is $\mathbf{s}_e = \hbar\bs{\sigma}/2$. Since for $\alpha>0$ the itinerant electrons interact anti-ferromagnetically with the impurity spin $\mathbf{S}\propto\mathbf{s}_e$, see Fig. \ref{Fig2}, the interaction is ferromagnetic in terms of the impurity magnetic moment $\bs{\mathcal{M}}\propto\bs{\mu}_e$.}. On the other hand, if impurities are made of rear earth elements, e.g. samarium Sm, with partially filled f-shells, the local exchange scattering is ferromagnetic, $\alpha < 0$. 
Finally, for the case of ferromagnetically aligned magnetic impurities, there is an extra term in the Hamiltonian, see Eq. (\ref{Himp_ferro}). The first term in Eq. (\ref{Himp_ferro}) describes a homogeneous background magnetic field created by impurity magnetic moments, see Fig. \ref{Fig1}(b). Here $\beta\sim 1$ is a dimensionless fitting parameter, which has a meaning of a geometrical structure factor determined by the actual impurity distribution in space, and $n$ is the density of impurities.
\subsection{Quasiclassical Green's function}
In our calculations we use the quasiclassical Green's function formalism \cite{Eilenberger1968,Larkin_Ovchinnikov_1969,Eschrig_PRB2000,Eschrig_PRB2009}. Since we aim at describing equilibrium properties of our system, the central object of interest is the Matsubara Green's function \cite{Matsubara1955} $\hat{g}(\epsilon_n,\mathbf{p}_{F},\mathbf{r})$. The ``hat" denotes a $4\times4$ matrix structure in combined spin and Nambu (or particle-hole) space. The Green's function depends on the Matsubara frequency $\epsilon_n = 2\pi k_BT(n+1/2)$, the quasiparticle momentum on the Fermi surface $\mathbf{p}_{F}$, and the spatical coordinate $\mathbf{r}$. Here $T$ is the temperature and $k_B$ is the Boltzmann constant. The propagator $\hat{g}$ satisfies the quasiclassical Eilenberger equation \cite{Eilenberger1968}
\begin{align}
\label{EilEqn}
[i\epsilon_n\hat{\tau}_3 - \hat{h}, \hat{g}] +i\hbar \mathbf{v}_{F}\cdot\boldsymbol{\nabla} \hat{g}
= \hat{0},
\end{align}
where $\hat{\tau}_3$ is the third Pauli matrix in Nambu space, $\mathbf{v}_{F}$ is the Fermi velocity, and $\hbar$ is the Planck constant. Equation (\ref{EilEqn}) has to be supplemented by a normalization condition $\hat{g}^2 = -\pi^2\hat{1}$. In Eq. (\ref{EilEqn}) the self-energy matrix $\hat{h}$ is parametrized as
\begin{align}
\hat{h}=
\begin{pmatrix}
\Sigma & \Delta \\
\tilde{\Delta} & \tilde{\Sigma}
\end{pmatrix},
\label{SEgeneral}
\end{align}
where each element has a $2\times2$ structure in spin space, and we have introduced the ``tilde''-operation defined as
\begin{align}
\tilde{Y}(\epsilon_n,\mathbf{p}_{\mathrm{F}},\mathbf{r})=Y(\epsilon_n,-\mathbf{p}_{\mathrm{F}},\mathbf{r})^{\ast}.\label{TildeOp}
\end{align}
\subsubsection{Riccati parametrization}
In order to solve Eq. (\ref{EilEqn}) it is convenient to employ the Riccati parametrization \cite{Schopohl_Maki_PRB1995,*Schopohl_1998,Eschrig_PRB2000,Eschrig_PRB2009} for the propagator $\hat{g}$. It is realized in terms of ``coherence functions" $\gamma$ and $\tilde{\gamma}$ as \cite{Eschrig_PRB2009,Grein_PRB2013}
\begin{align}
\begin{array}{c}
\hat{g}=\mp 2\pi i
\begin{pmatrix}
\mathcal{G} & \mathcal{F} \\
-\tilde{\mathcal{F}} & -\tilde{\mathcal{G}}
\end{pmatrix}
\pm i\pi\hat{\tau}_3,\\[1.0em]
\mathcal{G} = (1-\gamma\tilde{\gamma})^{-1},\;\;\mathcal{F} = \mathcal{G}\gamma,
\end{array}
\label{GRiccati}
\end{align}
where $\mp$ and $\pm$ correspond to positive and negative Matsubara frequencies, respectively. This allows us to rewrite Eq. (\ref{EilEqn}) as a system of (Riccati-type) transport equations for coherence functions,
\begin{align}
\begin{array}{l}
(i\hbar \mathbf{v}_{F}\cdot\boldsymbol{\nabla}+2i\epsilon_n)\gamma = \gamma\tilde{\Delta}\gamma + \Sigma\gamma - \gamma\tilde{\Sigma} - \Delta,\\ [1.0em]
(i\hbar \mathbf{v}_{F}\cdot\boldsymbol{\nabla}-2i\epsilon_n)\tilde{\gamma} = \tilde{\gamma}\Delta\tilde{\gamma} + \tilde{\Sigma}\tilde{\gamma} - \tilde{\gamma}\Sigma - \tilde{\Delta}.
\end{array}
\label{RiccatiEqs}
\end{align}
The coherence functions for negative and positive Matsubara frequencies are related via a symmetry relation
\begin{align}
\gamma(\epsilon_n < 0)\vert_{\epsilon_n \rightarrow -\epsilon_n} = [\tilde \gamma(\epsilon_n > 0)]^\dagger,
\end{align}
which allows us to express all physical observables in terms of only positive Matsubara frequencies. Therefore we consider $\gamma$ and $\tilde{\gamma}$ only for $\epsilon_n > 0$ below.
\subsubsection{Analytic continuation: retarded and advanced propagators}
Most properties of a system in thermal equilibrium can be calculated directly from the Matsubara propagator $\hat{g}$. However, energy-resolved quantities like, for example, density of states need to be computed on the real energy axis. So, we need a recipe how to perform an analytic continuation from Matsubara frequencies to real energies. In most common cases analytical continuation is a formal procedure, which establishes a one-to-one correspondence between the Matsubara $\hat{g}$ and the retarded $\hat{g}^R$ and advanced $\hat{g}^A$ propagators, which reads \cite{Baym1961,Serene_Rainer_PhysRep1983}
\begin{align}
\begin{array}{l}
\hat{g}^R(\epsilon) = \hat{g}(\epsilon_n > 0)\vert_{i\epsilon_n\rightarrow\epsilon + i0^{+}},\\[1.0em]
\hat{g}^A(\epsilon) = \hat{g}(\epsilon_n < 0)\vert_{i\epsilon_n\rightarrow\epsilon - i0^{+}}.
\end{array}
\label{AnalCont}
\end{align}

For real energies the ``tilde"-operation is defined as
\begin{align}
\tilde{Y}(\epsilon,\mathbf{p}_{\mathrm{F}},\mathbf{r})=Y(-\epsilon,-\mathbf{p}_{\mathrm{F}},\mathbf{r})^{\ast}.\label{TildeOp_re}
\end{align}

\subsection{Self-consistency equations and observables}
Since the formalism we use is a mean field theory we need to supplement the above equations with the corresponding self-consistency equations for the self-energies. Because we have no control over the distribution of impurities, we have to make several standard yet important approximations. In what follows we describe properties of the system assuming averaging over different impurity configurations \footnote{In order for these results to be applicable to a given sample with a given distribution of impurities, the size of the system must be much larger than the phase coherence length. Then, the system is said to be self-averaging. For the case of small superconducting islands one has to consider the actual spatial arrangement of impurities to make sensible predictions.}. Furthermore we assume a dilute concentration of impurities which allows us to use the so-called non-crossing t-matrix approximation for finding the impurity self-energy \cite{Rammer_book1998}. The last important assumption, which is usually implicitly used, is that the superconducting order parameter $\Delta_0$ (assumed real in this paper) is a self-averaging quantity which means that in presence of a dilute impurity distribution the order parameter is homogeneous on the superconducting coherence length scale \cite{Bennemann_Ketterson_2008}. After impurity averaging the system becomes homogeneous and thus Eq. (\ref{RiccatiEqs}) transforms into a system of algebraic equations.

We introduce for the rest of the paper a set of parameters characterizing the impurity subsystem,
\begin{align}
u_{0,\mathrm{S}} = \pi N_F v_{0,\mathrm{S}},\;\; \Gamma = \frac{n}{\pi N_F},\label{u_Gamma_def}
\end{align} 
where $N_F$ is the density of states per spin and unit volume at the Fermi energy in the normal state. Below we summarize all the self-consistency equations as well as observables considered in this paper for the two models of magnetic impurities mentioned above.

\subsubsection{Randomly oriented impurities}

We have two types of self-consistency equations: (i) for the impurity self-energy and (ii) for the order parameter $\Delta_0$. For the case of unpolarized impurities, see Fig. \ref{Fig1}(a), the coherence functions as well as the superconducting pairing self-energy have spin-singlet structure
\begin{align}
\gamma = i\sigma_2\gamma_0,\;\;\tilde{\gamma} = i\sigma_2\tilde{\gamma}_0,\;\;\Delta_S = i\sigma_2\Delta_0.
\label{gamma_structure_rand}
\end{align}
Then the impurity self-energy in the t-matrix approximation can be written as (see Appendix \ref{App_t_matr})
\begin{align}
&\Sigma_{\mathrm{imp}}(\epsilon_n) = \frac{\Gamma}{d}\Bigl\{u_0(1+u_0^2-\alpha^2u_{\mathrm{S}}^2) -i\left[u_0^2+\alpha^2u_{\mathrm{S}}^2\right. \notag\\
&\phantom{\Sigma_{\mathrm{imp}}(\epsilon_n) }\left.+(u_0^2-\alpha^2u_{\mathrm{S}}^2)^2\right](2\mathcal{G}-1)
\Bigr\},\notag\\
&\Delta_{\mathrm{imp}}(\epsilon_n) = i\sigma_2\Delta_{0,\mathrm{imp}}(\epsilon_n),\label{SE_imps_rand}\\
&\Delta_{0,\mathrm{imp}}(\epsilon_n) =  -\frac{2i\Gamma}{d}\left[u_0^2-\alpha^2u_{\mathrm{S}}^2+(u_0^2-\alpha^2u_{\mathrm{S}}^2)^2\right]\mathcal{G}\gamma_0,\notag\\
& d = (1+u_0^2-\alpha^2u_{\mathrm{S}}^2)^2+4\alpha^2u_{\mathrm{S}}^2(2\mathcal{G}-1)^2. \notag
\end{align}
The corresponding ``tilded" counterparts $\tilde{\Sigma}_{\mathrm{imp}}$ and $\tilde{\Delta}_{\mathrm{imp}}$ are obtained by applying the tilde-operation, see Eq. (\ref{TildeOp}). It is convenient to introduce two new quantities
\begin{align}
E_n = \epsilon_n + i\frac{\Sigma_{\mathrm{imp}}-\tilde{\Sigma}_{\mathrm{imp}}}{2},\;\;D_n = \Delta_0 + \Delta_{0,\mathrm{imp}}.
\end{align}
It can be shown that in terms of the new quantities the solution to the (homogeneous) Riccati equations reads
\begin{align}
\gamma_0(\epsilon_n) = i\frac{D_n}{E_n+\sqrt{D_n^2+E_n^2}},\;\;\tilde{\gamma}_0(\epsilon_n) = -\gamma_0(\epsilon_n),
\label{gamma_0_rand}
\end{align}
and the symmetry relations $\tilde{E}_n = E_n$ and $\tilde{D}_n = D_n$ hold. Using equations (\ref{SE_imps_rand})-(\ref{gamma_0_rand}) we can write down a pair of self-consistency equations for $E_n$ and $D_n$,
\begin{align}
& E_n \!= \epsilon_n + \Gamma\frac{\left[u_0^2+\alpha^2u_{\mathrm{S}}^2+(u_0^2-\alpha^2u_{\mathrm{S}}^2)^2\right]\!E_n\sqrt{D_n^2+E_n^2}}{\mathcal{D}_n},\notag\\
& D_n \!= \Delta_0 + \Gamma\frac{\left[u_0^2-\alpha^2u_{\mathrm{S}}^2+(u_0^2-\alpha^2u_{\mathrm{S}}^2)^2\right]\!D_n\sqrt{D_n^2+E_n^2}}{\mathcal{D}_n},\notag\\
&\mathcal{D}_n = (1+u_0^2-\alpha^2u_{\mathrm{S}}^2)^2(D_n^2+E_n^2)+4\alpha^2u_{\mathrm{S}}^2E_n^2.
\label{SE_imps_rand_2}
\end{align}
Let us now find the self-consistency equation for the order parameter. By definition,
\begin{align}
\Delta_0 = \frac{\lambda N_F}{2}k_BT\sum_{|\epsilon_n|<\epsilon_c}\int\frac{d\Omega_{\mathbf{p}_{F}}}{4\pi}\mathrm{Tr}[i\sigma_2f(\epsilon_n,\mathbf{p}_{F})],
\label{OP_gen}
\end{align}
where $f(\epsilon_n,\mathbf{p}_{F})=\mp2\pi i\mathcal{F}$ for positive and negative Matsubara frequencies, respectively [see Eq. (\ref{GRiccati})]. $\lambda < 0$ is the electron-phonon coupling constant and $\epsilon_c$ is the high-energy cut-off of the order of the Debye frequency. Equation (\ref{OP_gen}) in our case simplifies to
\begin{align}
\Delta_0\ln\frac{T}{T_{c0}} = 2\pi k_B T\sum_{\epsilon_n>0}\mathrm{Re}\left[\frac{D_n}{\sqrt{D_n^2+E_n^2}}-\frac{\Delta_0}{\epsilon_n}\right],
\label{OP_rand}
\end{align}
where $T_{c0}$ is the clean limit superconducting transition temperature. In Eq. (\ref{OP_rand}) we have eliminated the coupling constant $\lambda$ in favor of $T_{c0}$ by using the standard relation \cite{Serene_Rainer_PhysRep1983,Grein_PRB2013}
\begin{align}
\frac{1}{|\lambda|N_F} = \ln\frac{T}{T_{c0}} + 2\pi k_BT\sum_{0<\epsilon_n<\epsilon_c}\frac{1}{\epsilon_n}.
\end{align}
Besides the magnitude of the order parameter $\Delta_0$ it is useful to know the actual superconducting transition temperature $T_c$ of the disordered system. Linearizing Eq. (\ref{OP_rand}) with respect to $\Delta_0$ we obtain
\begin{align}
&\ln\frac{T_c}{T_{c0}} = \psi\left(\frac{1}{2}\right) - \psi\left(\frac{1}{2}+\frac{\Gamma_{\mathrm{eff}}}{2\pi k_B T_c}\right),\label{Tc_eqn_rand}\\
& \Gamma_{\mathrm{eff}} = 2\alpha^2u_{\mathrm{S}}^2\Gamma/d_c,\;\;d_c = (1+u_0^2-\alpha^2u_{\mathrm{S}}^2)^2+4\alpha^2u_{\mathrm{S}}^2, \notag
\end{align}
where $\psi(x)$ is the digamma function. Equation (\ref{Tc_eqn_rand}) is similar to the famous Abrikosov-Gor'kov formula \cite{Abrikosov_Gorkov_1961}, with an effective pair-breaking parameter $\Gamma_{\mathrm{eff}}$.

As soon as the self-consistent solution for $\Delta_0$ is found we can compute the density of states by using the recipe given in Eq. (\ref{AnalCont}). The coherence function becomes,
\begin{align}
\gamma_0(\epsilon) = -\frac{D}{E+i\sqrt{D^2-E^2}},\label{gamma_0_retarded}
\end{align}
while the final result for the density of states reads
\begin{align}
N(\epsilon) = 2N_F\mathrm{Re}\left[\frac{1+\gamma_0^2}{1-\gamma_0^2}\right] = 2N_F\mathrm{Im}\left[\frac{E}{\sqrt{D^2-E^2}}\right],
\label{DOS_rand_eqn}
\end{align}
where $E$ and $D$ are the real-energy versions of $E_n$ and $D_n$ defined as
\begin{align}
iE_n(\epsilon_n)\rightarrow E(\epsilon),\;\;D_n(\epsilon_n)\rightarrow D(\epsilon).
\end{align}
They satisfy
\begin{align}
& E \!= \epsilon + \Gamma\frac{\left[u_0^2+\alpha^2u_{\mathrm{S}}^2+(u_0^2-\alpha^2u_{\mathrm{S}}^2)^2\right]\!E\sqrt{D^2-E^2}}{\mathcal{D}},\notag\\
& D \!= \Delta_0 + \Gamma\frac{\left[u_0^2-\alpha^2u_{\mathrm{S}}^2+(u_0^2-\alpha^2u_{\mathrm{S}}^2)^2\right]\!D\sqrt{D^2-E^2}}{\mathcal{D}},\notag\\
&\mathcal{D} = (1+u_0^2-\alpha^2u_{\mathrm{S}}^2)^2(D^2-E^2)-4\alpha^2u_{\mathrm{S}}^2E^2.
\end{align}
\subsubsection{Ferromagnetically ordered impurities}
For the case of ferromagnetically ordered impurities, see Fig. \ref{Fig1}(b), there is an exchange field in the system, which makes properties of opposite spin quasiparticles inequivalent. Thus it is convenient to parametrize the coherence functions and impurity self-energies as
\begin{align}
&\gamma = 
\begin{pmatrix}
0 & \gamma_{\uparrow} \\
-\gamma_{\downarrow} & 0
\end{pmatrix},\;\;
\tilde{\gamma} = 
\begin{pmatrix}
0 & \tilde{\gamma}_{\uparrow} \\
-\tilde{\gamma}_{\downarrow} & 0
\end{pmatrix},\notag\\
&\Sigma_{\mathrm{imp}} = 
\begin{pmatrix}
\Sigma_{\uparrow} & 0\\
0 & \Sigma_{\downarrow}
\end{pmatrix},\;\;
\Delta_{\mathrm{imp}} = 
\begin{pmatrix}
0 & \Delta_{\uparrow} \\
-\Delta_{\downarrow} & 0
\end{pmatrix},\;\;
\label{gamma_structure_ferro}
\end{align}
which would simplify to the randomly oriented case considered before upon setting the opposite spin-components equal. Now we introduce
\begin{align}
\epsilon_{n\uparrow} = \epsilon_n + i\beta\Gamma u_{\mathrm{S}},\;\;
\epsilon_{n\downarrow} = \epsilon_n - i\beta\Gamma u_{\mathrm{S}},
\end{align}
and solve the t-matrix equation (see Appendix \ref{App_t_matr}), by analogy with the previous case. Then we introduce the spin-dependent self-energies
\begin{align}
\begin{array}{l}
\displaystyle E_{n\uparrow} = \epsilon_{n\uparrow} + i\frac{\Sigma_{\uparrow}-\tilde{\Sigma}_{\downarrow}}{2},\;\;E_{n\downarrow} = \epsilon_{n\downarrow} + i\frac{\Sigma_{\downarrow}-\tilde{\Sigma}_{\uparrow}}{2},\\[1em]
D_{n\uparrow} = \Delta_0 + \Delta_{\uparrow},\;\;D_{n\downarrow} = \Delta_0 + \Delta_{\downarrow},
\end{array}
\end{align}
in terms of which the solution to the homogeneous Riccati equations is given by
\begin{align}
&\gamma_{\uparrow}(\epsilon_n) = i\frac{D_{n\uparrow}}{E_{n\uparrow}+\sqrt{D_{n\uparrow}^2+E_{n\uparrow}^2}},\notag\\
&\gamma_{\downarrow}(\epsilon_n) = i\frac{D_{n\downarrow}}{E_{n\downarrow}+\sqrt{D_{n\downarrow}^2+E_{n\downarrow}^2}},\\
&\tilde{\gamma}_{\uparrow}(\epsilon_n) = -\gamma_{\downarrow}(\epsilon_n),\;\;\tilde{\gamma}_{\downarrow}(\epsilon_n) = -\gamma_{\uparrow}(\epsilon_n),\notag
\end{align}
and the following symmetries hold
\begin{align}
\tilde{E}_{n\chi} = E_{n\chi},\;\;\tilde{D}_{n\chi} = D_{n\chi},\;\;\chi = \left\{\uparrow,\downarrow\right\}.
\end{align}
Spin-dependent self-energies satisfy the following self-consistency equations
\begin{align}
& E_{n\chi} = \epsilon_{n\chi} + \Gamma\frac{(u_0^2-\alpha^2u_{\mathrm{S}}^2)E_{n\chi}\pm i\alpha u_{\mathrm{S}}\sqrt{D_{n\chi}^2+E_{n\chi}^2}}{\mathcal{D}_{n\chi}},\notag\\
& D_{n\chi} = \Delta_0 + \Gamma\frac{(u_0^2-\alpha^2u_{\mathrm{S}}^2)D_{n\chi}}{\mathcal{D}_{n\chi}},\;\;\chi = \left\{\uparrow,\downarrow\right\},\label{SE_eqn_ord}\\
&\mathcal{D}_{n\chi} = (1+u_0^2-\alpha^2u_{\mathrm{S}}^2)\sqrt{D_{n\chi}^2+E_{n\chi}^2}\pm 2i\alpha u_{\mathrm{S}}E_{n\chi},\notag
\end{align}
where the upper (lower) sign refers to spin-up (spin-down) quasiparticles.
Next, we find the self-consistency equation for the order parameter, which is still given by Eq. (\ref{OP_gen}). In this case it reads
\begin{align}
\Delta_0\ln\frac{T}{T_{c0}} = 2\pi k_B T&\sum_{\epsilon_n>0}\mathrm{Re}\left[\frac{D_{n\uparrow}}{2\sqrt{D_{n\uparrow}^2+E_{n\uparrow}^2}} \right.\notag\\
&\left.+ \frac{D_{n\downarrow}}{2\sqrt{D_{n\downarrow}^2+E_{n\downarrow}^2}} -\frac{\Delta_0}{\epsilon_n}\right].
\label{OP_ferro}
\end{align}
Linearizing this equation with respect to $\Delta_0$ we can also find the equation for the critical temperature $T_c$,
\begin{align}
&\ln\frac{T_c}{T_{c0}} = \frac{1}{2}\sum_{n=0}^{\infty}\Biggl\{\notag\\
&\phantom{\ln\frac{T_c}{T_{c0}} =}\left[n+\frac{1}{2}+i\frac{\Gamma}{2\pi k_B T_c}\frac{u_{\mathrm{S}}(\alpha+\beta d_{c\uparrow})}{d_{c\uparrow}}\right]^{-1}
\notag\\
&\phantom{\ln\frac{T_c}{T_{c0}} }+\left[n+\frac{1}{2}-i\frac{\Gamma}{2\pi k_B T_c}\frac{u_{\mathrm{S}}(\alpha+\beta d_{c\downarrow})}{d_{c\downarrow}}\right]^{-1}\notag\\
&\phantom{\ln\frac{T_c}{T_{c0}} }-2\left(n+\frac{1}{2}\right)^{-1}
\Biggr\},\label{Tc_eqn_ferro}\\
& d_{c\uparrow} = 1+u_0^2-\alpha^2u_{\mathrm{S}}^2 + 2i\alpha u_{\mathrm{S}},\;\; d_{c\downarrow} = d_{c\uparrow}^{\ast}. \notag
\end{align}
Equation (\ref{Tc_eqn_ferro}) could in principle be rewritten in terms of digamma functions of complex argument, but the form above was used in the actual calculations.

The (spin-resolved) density of states is computed via analytical continuation to the real energy axis, see Eq. (\ref{AnalCont}), by analogy with the previous case, and the final expression reads
\begin{align}
N_{\uparrow,\downarrow}(\epsilon) = N_F\mathrm{Im}\left[\frac{E_{\uparrow,\downarrow}}{\sqrt{D_{\uparrow,\downarrow}^2-E_{\uparrow,\downarrow}^2}}\right].
\end{align}
Here $E_{\uparrow,\downarrow}$ and $D_{\uparrow,\downarrow}$ satisfy
\begin{align}
& E_{\chi} = \epsilon_{\chi} + \Gamma\frac{(u_0^2-\alpha^2u_{\mathrm{S}}^2)E_{\chi} \mp \alpha u_{\mathrm{S}}\sqrt{D_{\chi}^2-E_{\chi}^2}}{\mathcal{D}_{\chi}},\notag\\
& D_{\chi} = \Delta_0 + \Gamma\frac{(u_0^2-\alpha^2u_{\mathrm{S}}^2)D_{\chi}}{\mathcal{D}_{\chi}},\;\;\chi = \left\{\uparrow,\downarrow\right\},\\
&\mathcal{D}_{\chi} = (1+u_0^2-\alpha^2u_{\mathrm{S}}^2)\sqrt{D_{\chi}^2-E_{\chi}^2}\pm 2\alpha u_{\mathrm{S}}E_{\chi}\notag,
\end{align}
where $\epsilon_{\uparrow,\downarrow} = \epsilon \mp \beta\Gamma u_{\mathrm{S}}$.

Another observable we are interested in for the case of polarized impurities is the magnetization in the system. In the framework of our model, the expression for magnetization reads
\footnote{A crucial feature of the quasiclassical theory is that there exists a separation of energy (or length) scales, e.g. $\Delta_0\ll E_F$, where $E_F$ is the Fermi energy. This enables one to introduce a small parameter $\mathtt{small}$, which is used as an expansion parameter for the full microscopic propagators and self-energies \cite{Serene_Rainer_PhysRep1983,Sauls_FLT1994}. In our case, apart from the order parameter being small compared to $E_F$ we have to assume that $nv_{0,\mathrm{S}}\ll E_F$. Following the procedure of calculating physical observables described in Ref. [\onlinecite{Serene_Rainer_PhysRep1983}] we have obtained Eq. (\ref{M_eqn}). The first term is the so-called high-energy correction, which is not captured by quasiclassics and has to be computed separately.},
\begin{align}
&\phantom{\mathrm{M} = 2(\alpha+\beta)\Gamma u_{\mathrm{S}}}\mathbf{M} = \mathrm{M}\mathbf{m},\notag\\
&\mathrm{M} = 2(\alpha+\beta)\Gamma u_{\mathrm{S}}\mu_BN_F - 2\pi k_B T\mu_BN_F \label{M_eqn}\\
&\times\sum_{\epsilon_n>0}\mathrm{Im}\left[\frac{E_{n\uparrow}}{\sqrt{D_{n\uparrow}^2+E_{n\uparrow}^2}}
-\frac{E_{n\downarrow}}{\sqrt{D_{n\downarrow}^2+E_{n\downarrow}^2}}\right].\notag
\end{align}
The first term corresponds to the normal state contribution, while the second one is the low-energy correction due to superconductivity. Note that for unpolarized impurities the self-energies are spin-degenerate and the magnetization vanishes (the first term vanishes after averaging over impurity directions, see Appendix \ref{App_t_matr}).

Finally, if Eq. (\ref{OP_ferro}) has more than one solution it is necessary to consider the difference between Gibbs free energies in the superconducting and normal states \cite{Eilenberger1968,Serene_Rainer_PhysRep1983,Keller1988,Sauls_FLT1994} in order to determine the physically relevant one. In our case the expression for the free energy difference is given by (see Appendix \ref{App_FE})
\begin{widetext}
\begin{align}
&\delta\Omega = \Delta_0^2N_F\ln\frac{T}{T_{c0}}+2\pi N_Fk_BT\sum_{\epsilon_n>0}\mathrm{Re}\left[\frac{\Delta_0^2}{\epsilon_n}-\frac{D_{n\uparrow}^2}{E_{n\uparrow}+\sqrt{D_{n\uparrow}^2+E_{n\uparrow}^2}}-\frac{D_{n\downarrow}^2}{E_{n\downarrow}+\sqrt{D_{n\downarrow}^2+E_{n\downarrow}^2}}\right]\notag\\
&+2\pi N_Fk_BT\sum_{\epsilon_n>0}\mathrm{Re}\left[\frac{(E_{n\uparrow}-\epsilon_{n\uparrow})E_{n\uparrow}+(D_{n\uparrow}-\Delta_0)D_{n\uparrow}}{\sqrt{D_{n\uparrow}^2+E_{n\uparrow}^2}}+\frac{(E_{n\downarrow}-\epsilon_{n\downarrow})E_{n\downarrow}+(D_{n\downarrow}-\Delta_0)D_{n\downarrow}}{\sqrt{D_{n\downarrow}^2+E_{n\downarrow}^2}}-E_{n\uparrow}-E_{n\downarrow}+2\epsilon_n\right]\notag\\
&-2\pi N_Fk_BT\sum_{\epsilon_n>0}\frac{\Gamma}{2}\mathrm{Re}\left\{\ln\left[\left(1+u_0^2-\alpha^2u_{\mathrm{S}}^2+\frac{2i\alpha u_{\mathrm{S}}E_{n\uparrow}}{\sqrt{D_{n\uparrow}^2+E_{n\uparrow}^2}}\right)\left(1+u_0^2-\alpha^2u_{\mathrm{S}}^2-\frac{2i\alpha u_{\mathrm{S}}E_{n\downarrow}}{\sqrt{D_{n\downarrow}^2+E_{n\downarrow}^2}}\right)\right]\right.\notag\\
&\left.-\ln\left[\left(1+u_0^2-\alpha^2u_{\mathrm{S}}^2\right)^2+4\alpha^2u_{\mathrm{S}}^2\right]\vphantom{\frac{E_{n\uparrow}}{\sqrt{E_{n\uparrow}^2}}}\!\right\}.
\label{FE_eqn}
\end{align}
\end{widetext}
The free energy difference for the unpolarized case is obtained by equating the opposite-spin self-energies (let $\beta=0$). The self-energies then satisfy Eq. (\ref{SE_imps_rand_2}).

\section{Results}\label{Sec_Res}
In this section we use the formulas obtained above to investigate properties of superconductors with magnetic impurities. For each observable we compare results of self-consistent numerical calculations for the cases of (i) randomly oriented impurity magnetic moments, and (ii) ferromagnetically ordered magnetic moments. For the ordered case, all numerical results presented below were obtained by setting $\beta = 1$ and $|\alpha| = 0.1$ in order to illustrate the Physics. These parameters should be considered as fit parameter when comparing theory with experiment.

Properties of the system depend very weakly on the strength of the scalar part of the impurity potential $u_0$. As was demonstrated in Ref. [\onlinecite{Okabe1983}], if the superconducting order parameter is isotropic the scalar part $u_0$ enters the theory only through the energy of the bound state. This can be accounted for by introducing an effective exchange scattering amplitude $\left\{u_0,u_{\mathrm{S}}\right\}\rightarrow u_{\mathrm{S}}^{\mathrm{eff}}$. Therefore, in all of the results presented below we have taken $u_0=0$ to reduce the parameter space.

%As we show below, for ferromagnetically ordered impurities, it is important to consider the Gibbs free energy difference [see Eq. (\ref{FE_eqn})] for certain sets of parameters. This ensures that we filter out unphysical solutions to the self-consistency equation (\ref{OP_ferro}) for the order parameter. On the other hand, for unpolarized impurities, according to our numerical calculations the free energy difference is always negative whenever there is a non-zero solution to the order parameter equation (\ref{OP_rand}), so we do not show it.

%\FloatBarrier
\subsection{Superconducting transition temperature}
We start by computing the superconducting transition temperature as a function of impurity concentration. We note that there are three ways to do it. First, one can solve the linearized gap equation, see Eqs. (\ref{Tc_eqn_rand}) and (\ref{Tc_eqn_ferro}). Second, one can search for the value of temperature for which the full self-consistency equation [see Eqs. (\ref{OP_rand}) and (\ref{OP_ferro})] for the order parameter has zero solution. Third, one can find the temperature at which the free energy difference Eq. (\ref{FE_eqn}) changes sign. All methods give the same answer if the superconducting phase transition is second-order. When the phase transition is first-order, however, the only way to determine the physical transition temperature is by computing the free energy difference.
\begin{figure}[t]
	\includegraphics[width=\columnwidth]{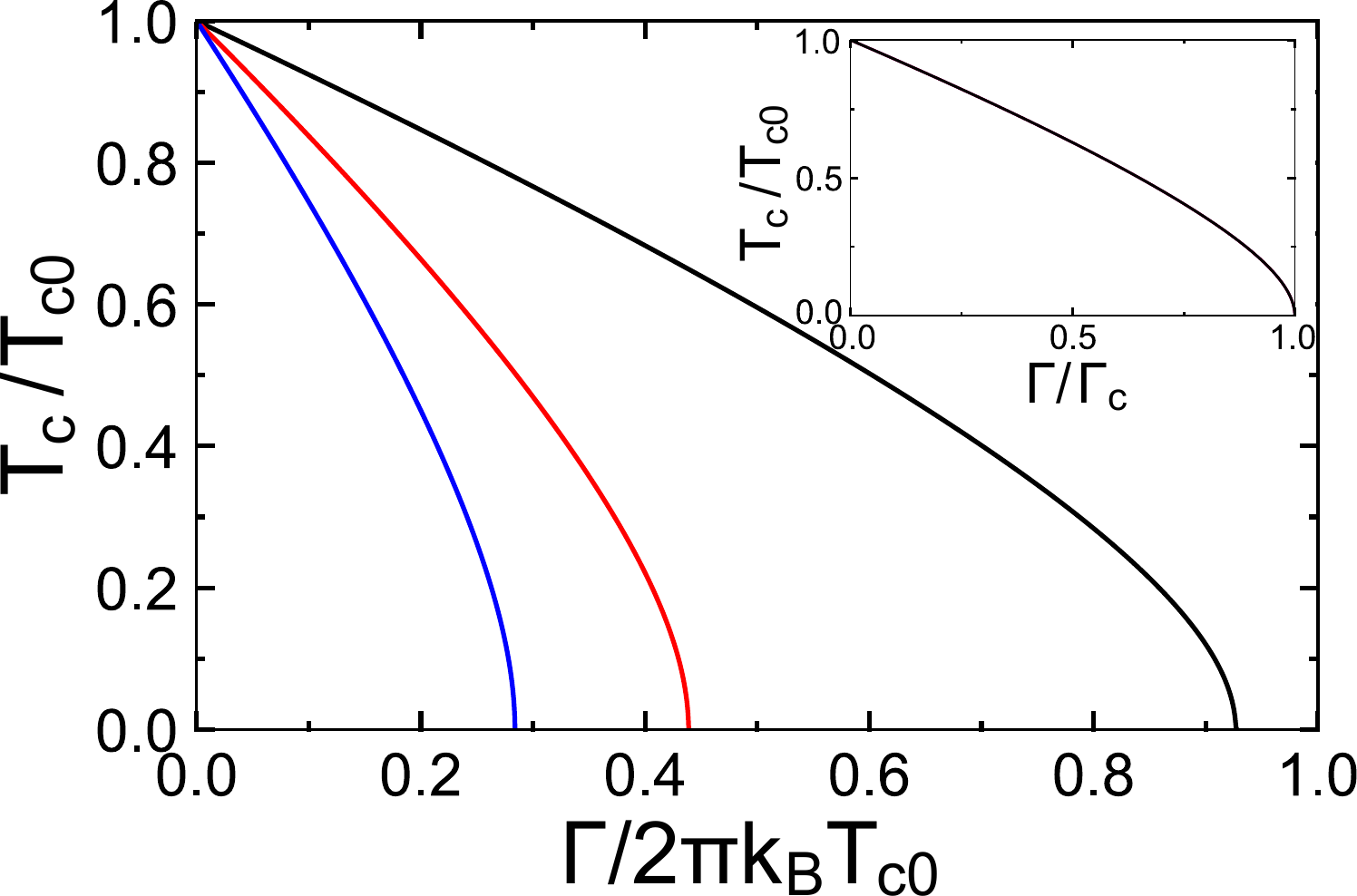}
	\caption{(Color online) Main plot: Transition temperature as a function of impurity density [see Eq. (\ref{u_Gamma_def})] for randomly oriented impurity spins. From right to left $u_{\mathrm{S}}=3,5,9$. Inset: The same data with the horizontal axis rescaled with respect to $\Gamma_c$, critical density at which $T_c = 0$ (see text).}
	\label{Tc_vs_Gamma_random}
\end{figure}
\subsubsection{Randomly oriented impurities}
For the case of randomly oriented impurity spins we observe suppression of the transition temperature which is similar to the classic result by Abrikosov and Gor'kov \cite{Abrikosov_Gorkov_1961}. The difference is that the effective pair-breaking parameter $\Gamma_{\mathrm{eff}}$ now depends on both the impurity concentration and the strength of the impurity potential, see Fig. \ref{Tc_vs_Gamma_random} and Eq. \eqref{Tc_eqn_rand}. Rescaling the horizontal axis of the figure with respect to $\Gamma_c/2\pi k_BT_{c0} = d_c/8\gamma\alpha^2u_{\mathrm{S}}^2$ (where $\gamma\approx1.78$ is the Euler constant), corresponding to the critical density of impurities at which $T_c = 0$, one can see that all the curves align, see inset of Fig. \ref{Tc_vs_Gamma_random}. The sign of $\alpha$ is unimportant for this case.
%to be $\Gamma/\Gamma_c$, where $\Gamma_c/2\pi k_BT_{c0} = d_c/8\gamma\alpha^2u_{\mathrm{S}}^2$ corresponds to the critical density at which $T_c = 0$, one can see that all the curves coincide with each other, see inset of Fig. \ref{Tc_vs_Gamma_random}. We note that the phase transition is always second order. The sign of $\alpha$ is unimportant for this case.
%Critical density of magnetic impurities is defined by $\Gamma^{\mathrm{cr}}_{\mathrm{eff}}/2\pi k_BT_{c0} = 1/4\gamma$ at which $T_c = 0$
\subsubsection{Ferromagnetically ordered impurities}
\begin{figure}[b]
	\includegraphics[width=\columnwidth]{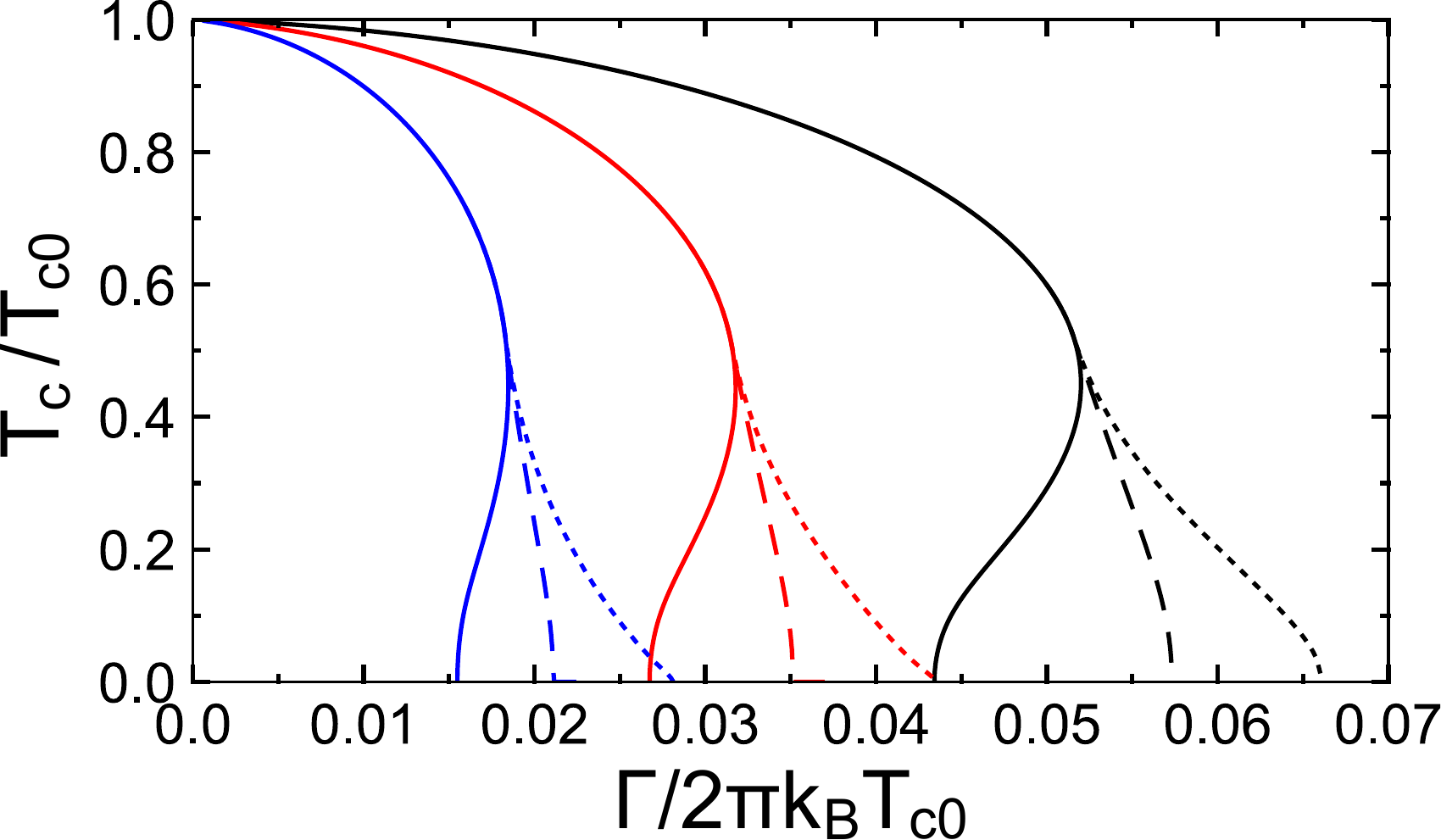}
	\caption{(Color online) Critical temperature as a function of impurity density for ferromagnetically ordered case and $\alpha>0$. From right to left we have $u_{\mathrm{S}}=3,5,9$. Solid lines are solutions to Eq. (\ref{Tc_eqn_ferro}). Dashed lines correspond to $\delta\Omega(T) = 0$, see Eq. (\ref{FE_eqn}). Dotted lines are found from Eq. (\ref{OP_ferro}) requiring $\Delta_0(T) = 0$.}
	\label{Tc_vs_Gamma_aligned_pos_a}
\end{figure}

Let us now discuss the case of ferromagnetically ordered impurities. In Fig. \ref{Tc_vs_Gamma_aligned_pos_a} we plot the critical temperature $T_c$ calculated by the three different methods described above. For small impurity concentrations all three curves coincide and the superconducting phase transition is second order.
For high enough concentrations we notice the appearance of the second solution, which means that there are two possible values of the order parameter in the system at low temperatures, as we will see below. The discrepancy between the three methods of finding $T_c$ is caused by the fact that the superconducting phase transition becomes first-order. This is expected since apart from the impurity scattering we also have a background Zeeman field in our model, see Eq. (\ref{Himp_ferro}). It is well known \cite{Clogston1962,Chandrasekhar1962,Gorkov_Rusinov_1964,Izyumov1974} that for high exchange fields the order of the superconducting phase transition changes to being first order. For an ordinary Zeeman term due to an external in-plan magnetic field, this is commonly known as the Clogston-Chandrasekhar limit. This means that the order parameter for temperatures close to $T_c$ is not small, and it is not allowed to linearize the self-consistency equation (\ref{OP_ferro}) to find $T_c$. The region in parameter space where this happens is the region where the solution to the linearized gap equation displays a back-bend (see solid lines in Fig. \ref{Tc_vs_Gamma_aligned_pos_a}). After solving the full (non-linear) self-consistency equation, we obtain the result depicted by the dotted lines in Fig. \ref{Tc_vs_Gamma_aligned_pos_a}. But, in order for the superconducting phase to exist it has to be more energetically favorable than the normal one, which can be checked by computing the Free energy difference. Searching for the temperature at which this condition fails, we obtain the physically correct solution depicted by the dashed lines in the figure. We have to mention that in our paper we assume a spatially constant and homogeneous order parameter, which corresponds to canonical Cooper pairing of electrons with equal and opposite momenta. However, there is a possibility of having a solution corresponding to Cooper pairs with finite center-of-mass momentum, which is known as the FFLO phase \cite{Fulde_Ferrel_1964,Larkin_Ovchinnikov_1964}. In our case this solution could exist, in principle, and it would lie in between the dashed and dotted lines in Fig. \ref{Tc_vs_Gamma_aligned_pos_a}. This is, however, out of scope of the present paper.

\begin{figure}[t]
	\includegraphics[width=\columnwidth]{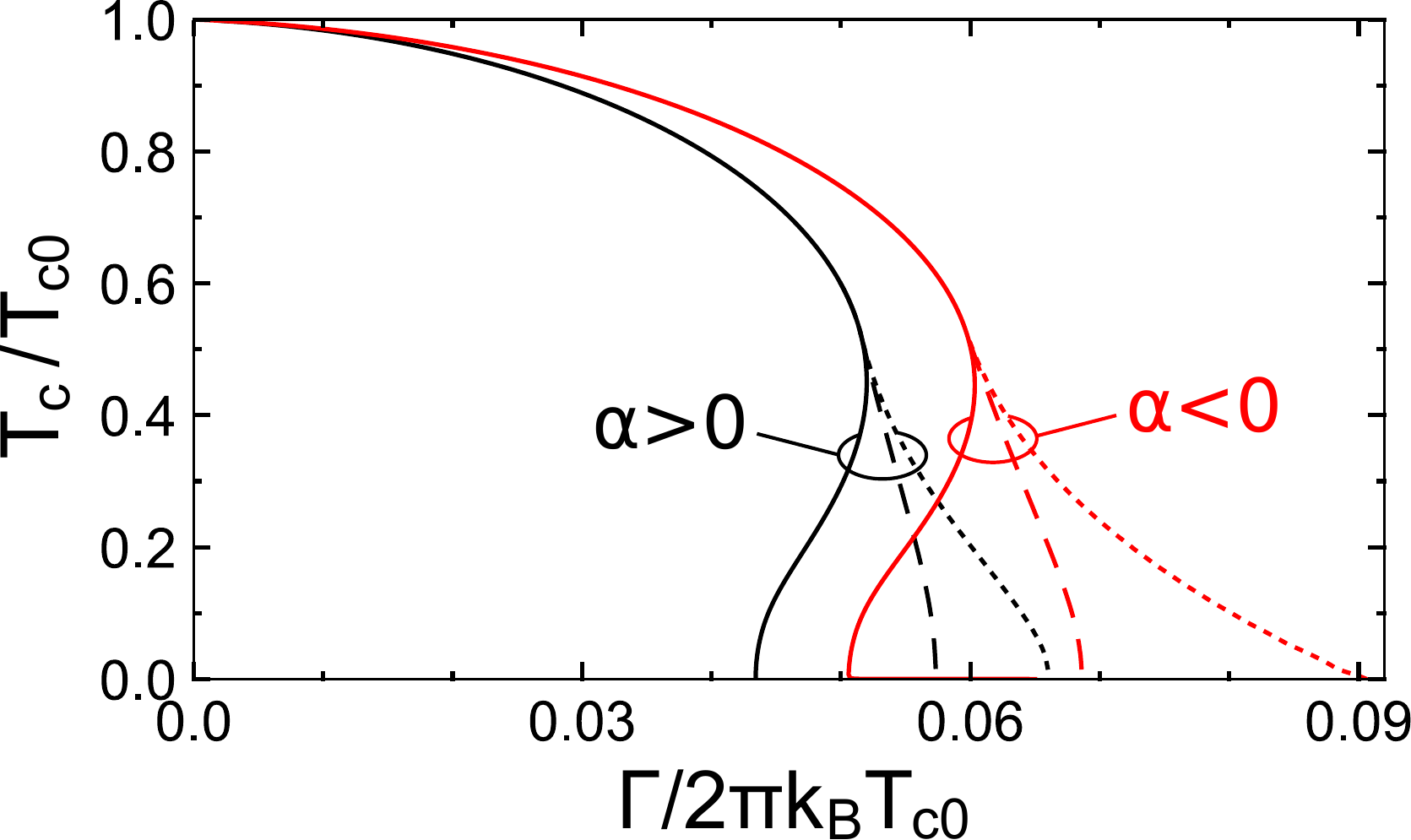}
	\caption{(Color online) Transition temperature as a function of impurity density for ferromagnetically ordered case with $u_{\mathrm{S}}=3$. The red (right) curve corresponds to $\alpha<0$, while the black (left) one is for $\alpha>0$, see Fig. \ref{Tc_vs_Gamma_aligned_pos_a}.}
	\label{Tc_vs_Gamma_aligned_neg_a}
\end{figure}
The results discussed above referred to the case of anti-ferromagnetic interaction of impurity spins with itinerant electrons, $\alpha>0$, see Fig. \ref{Fig2}. Let us now briefly discuss what happens if the local exchange interaction is ferromagnetic, $\alpha<0$. In Fig. \ref{Tc_vs_Gamma_aligned_neg_a} we compare the two cases and we can see that they look quite similar. For the parameters chosen in the figure, the critical temperature for $\alpha<0$ is always higher than for $\alpha>0$. In order to understand why it happens one has to remember that the suppression of the order parameter (and, as a consequence, of the critical temperature) as a function of impurity density is caused by the growing band of YSR states \cite{Shiba1968,Zittartz1972,Tsang1980,Bauriedl1981} shrinking the energy gap in the spectrum. Discussing the density of states below we will show that changing the sign of $\alpha$ changes the spin-polarization of the YSR states to the opposite one. Since we also have a Zeeman-like shift in our model, see Eq. (\ref{Himp_ferro}), which does not depend on the value of $\alpha$, the decrease of $T_c$ depends on how soon the impurity band meets the quasiparticle continuum \footnote{There is a small range of parameters where a gapless superconductivity \cite{Abrikosov_Gorkov_1961} is possible. Nevertheless, roughly speaking, transition to the normal state occurs when the impurity band fills in the energy gap.}. Therefore, one can find a set of parameters when, in contrast to the results in Fig. \ref{Tc_vs_Gamma_aligned_neg_a}, the critical temperature for $\alpha<0$ is smaller than for $\alpha>0$.

\subsection{Order parameter}
\subsubsection{Randomly oriented impurities}
\begin{figure}[t]
	\includegraphics[width=\columnwidth]{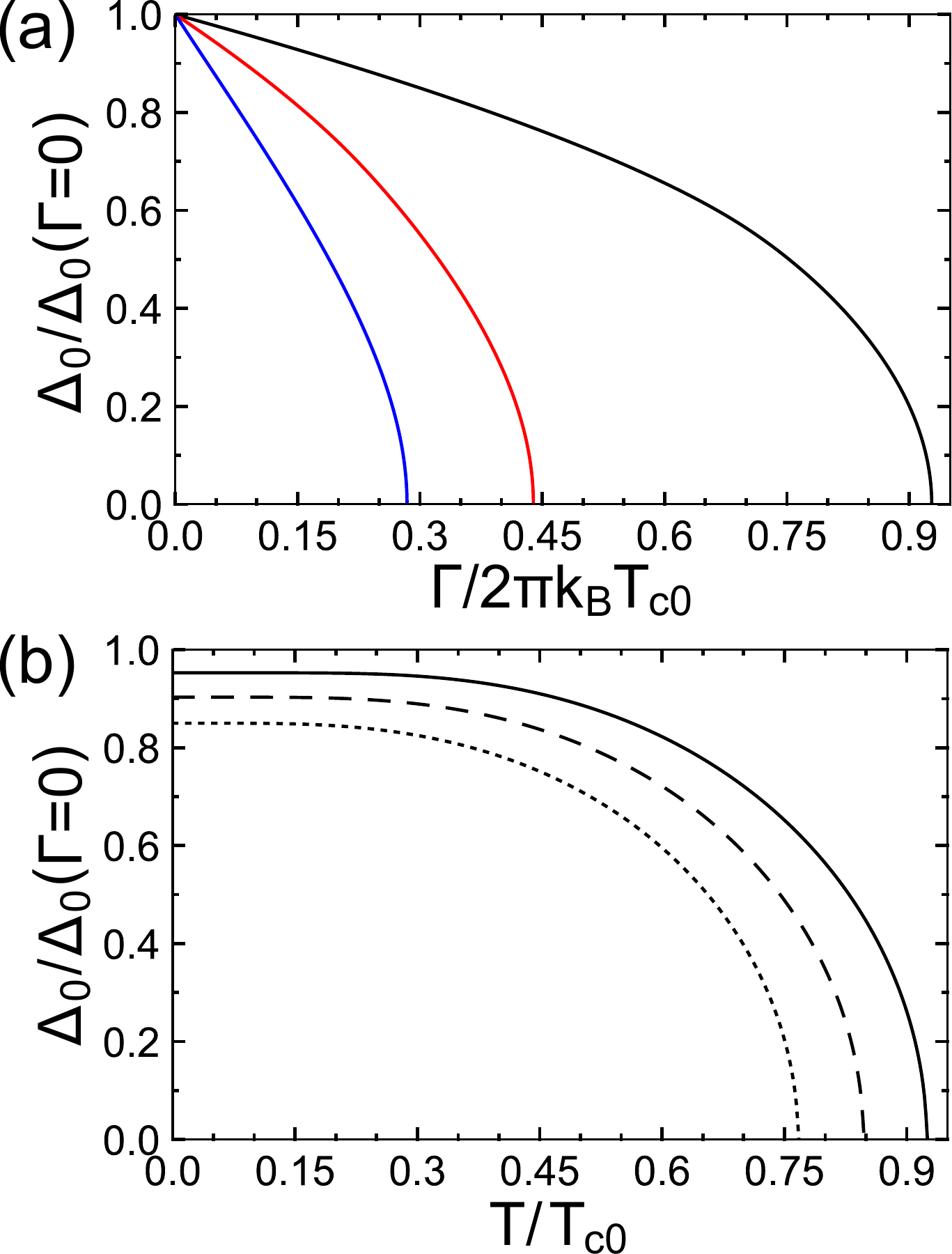}
	\caption{(Color online) Order parameter as a function of impurity density (a) and temperature (b) for randomly oriented impurity spins. (a): From right to left $u_s=3,5,9$ and $T=0.01\,T_{c0}$. (b): From top to bottom $\Gamma/2\pi k_B T_{c0}=0.1,0.2,0.3$ and $u_{\mathrm{S}}=3$.}
	\label{Delta_vs_Gamma_and_T_random}
\end{figure}
For randomly oriented impurity spins the order parameter behaves similarly to the Born limit considered by Abrikosov and Gor'kov \cite{Abrikosov_Gorkov_1961}. As can be seen in Fig. \ref{Delta_vs_Gamma_and_T_random}(a), the reduction of $\Delta_0$ as a function of impurity concentration is slower than the corresponding reduction of the critical temperature, see Fig. \ref{Tc_vs_Gamma_random}. Therefore if one plots $\Delta_0/k_BT_c$ as a function of density it seems that the order parameter grows \cite{Shevtsov_LT2014} (not shown in the figure). In Fig. \ref{Delta_vs_Gamma_and_T_random}(b) we plot the order parameter as a function of temperature for three different impurity densities. All the plots have similar shape and simply reflect the gradual reduction of $T_c$ and $\Delta_0$.

\begin{figure}[t]
\includegraphics[width=\columnwidth]{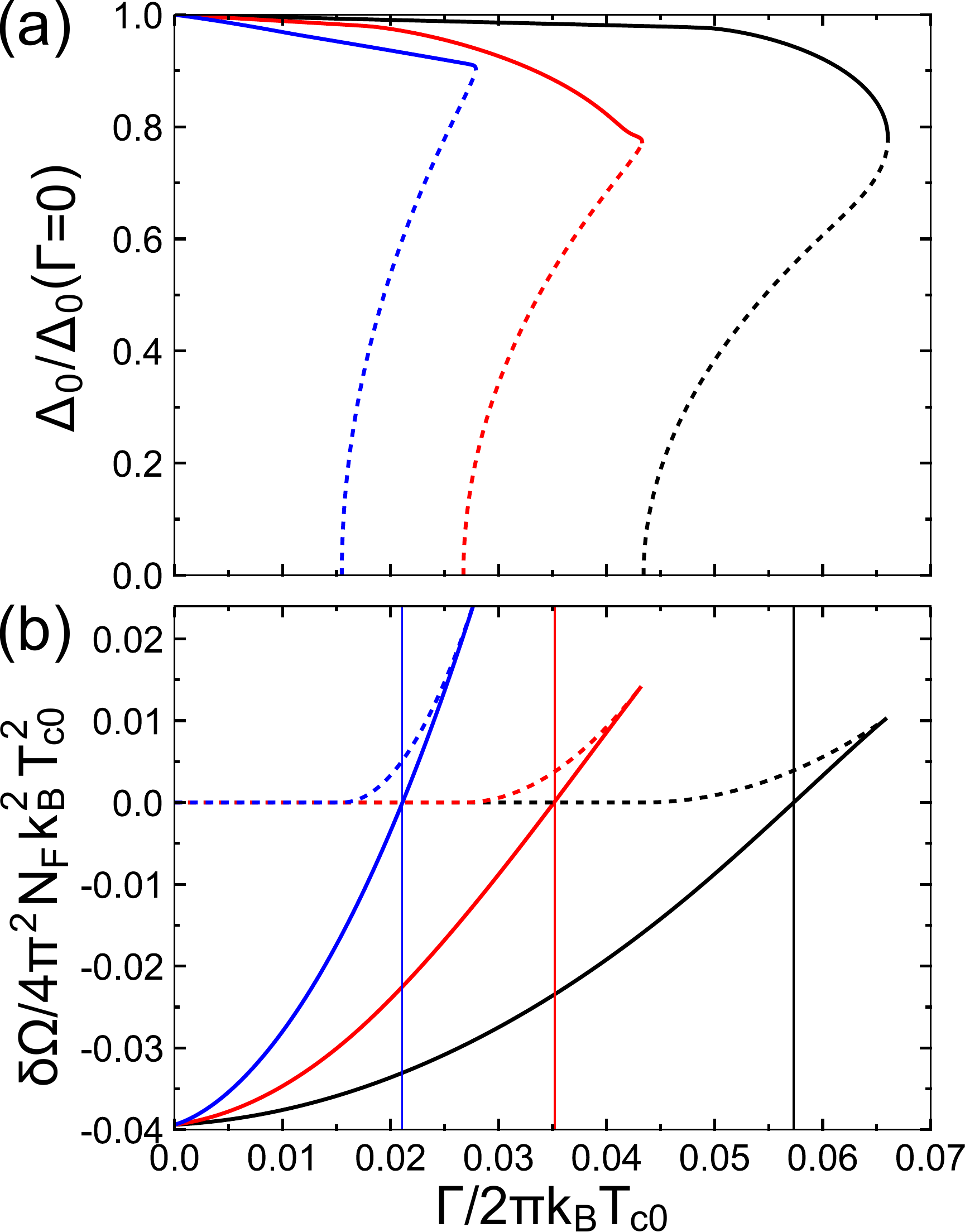}
\caption{(Color online) Order parameter (a) and the free energy difference (b) as a function of impurity density for ferromagnetically aligned impurity spins and $\alpha>0$. From right to left $u_s=3,5,9$ and $T=0.01\,T_{c0}$. Two different solutions to Eq. (\ref{OP_ferro}) and the corresponding free energy difference are plotted with solid and dotted lines. Thin vertical lines mark the density at which the free energy difference crosses zero.}
	\label{Delta_vs_Gamma_aligned_pos_a_FE}
\end{figure}

\begin{figure*}[t]
\includegraphics[width=\textwidth]{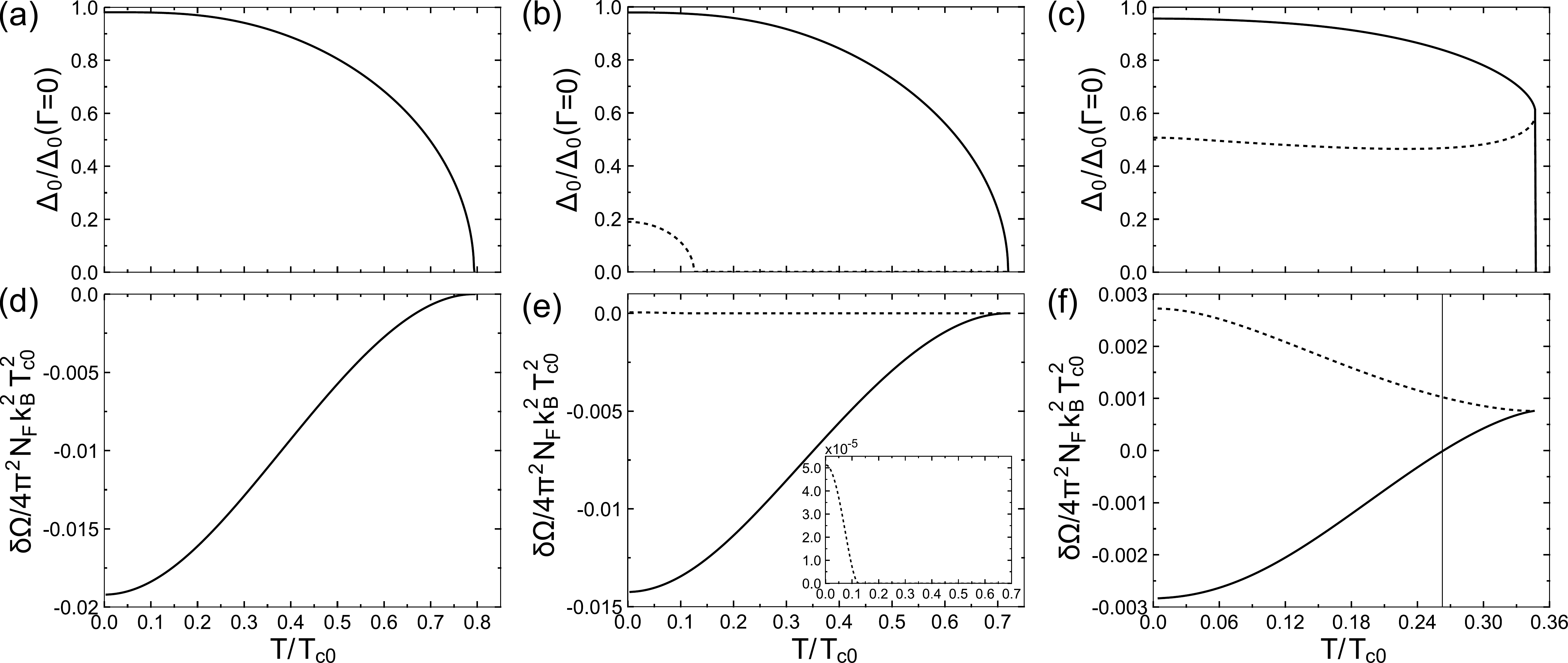}
\caption{Order parameter (a)-(c) and Gibbs free energy difference (d)-(f) as a function of temperature for ferromagnetically aligned impurity spins with $u_{\mathrm{S}}=3$ and $\Gamma/2\pi k_B T_{c0}$ equal to $0.04$ for (a) and (d), $0.045$ for (b) and (e), $0.055$ for (c) and (f). The vertical line in (f) depicts the temperature at which the free energy difference crosses zero. The inset in (d) shows a zoom on the dotted line in the main plot. The dotted lines in (b)-(c) depict the second solution to the order parameter equation (\ref{OP_ferro}) and the corresponding free energy difference in (e)-(f).}
	\label{Delta_vs_T_aligned_pos_a_FE}
\end{figure*}
\subsubsection{Ferromagnetically ordered impurities}
Let us now consider the case of ferromagnetically ordered impurity spins. In Fig. \ref{Delta_vs_Gamma_aligned_pos_a_FE}(a) we plot the order parameter as a function of impurity density for different values of $u_{\mathrm{S}}$. For low densities there is only one solution to Eq. (\ref{OP_ferro}), which decreases monotonically. However, for large enough densities the second solution emerges, depicted by the dotted lines in Fig. \ref{Delta_vs_Gamma_aligned_pos_a_FE}(a). In order to determine the physically relevant one we plot the difference of free energies in superconducting and normal states, see Fig. \ref{Delta_vs_Gamma_aligned_pos_a_FE}(b). As can be seen from the figure, the new solution has a non-negative free energy difference, which means that it is not energetically favorable and thus does not realize in practice. A similar conclusion was found in Ref. [\onlinecite{Izyumov1974}], where the superconductor to normal metal phase transition in a pure Zeeman field was analyzed. Another interesting point is that for even higher densities both solutions become energetically unfavorable and the system is no longer superconducting (in fact, it becomes a ferromagnet due to impurity ferromagnetism \cite{Abrikosov_Gorkov_1963}). The transition point at which it happens is depicted by thin vertical lines in Fig. \ref{Delta_vs_Gamma_aligned_pos_a_FE}(b).
If we change the sign of $\alpha$, corresponding to a different type of the local exchange interaction with impurities (see Fig. \ref{Fig2}), a similar behavior is observed, thus we do not show it here.
%\begin{figure}[t]
%\includegraphics[width=0.75\columnwidth]{plot_Delta_fn_Gamma_FE_ordered_neg_aV2.pdf}
%\caption{Order parameter (a) and the free energy difference (b) as a function of impurity density for ferromagnetically aligned impurity spins with $u_{\mathrm{S}}=3$. The red curve corresponds to $\alpha<0$, while the black one is for $\alpha>0$, see Fig. \ref{Delta_vs_Gamma_aligned_pos_a_FE}.}
%	\label{Delta_vs_Gamma_aligned_neg_a_FE}
%\end{figure}
%
%

To demonstrate the temperature dependence of the order parameter, we choose three different values of $\Gamma$. In Figs. \ref{Delta_vs_T_aligned_pos_a_FE}(a)-(c) we plot the order parameter as a function of temperature for $u_{\mathrm{S}}=3$ and $\Gamma = 0.04$, $0.045$ and $0.055$, respectively (see black lines in Figs. \ref{Tc_vs_Gamma_aligned_pos_a} and \ref{Delta_vs_Gamma_aligned_pos_a_FE}). In Figs. \ref{Delta_vs_T_aligned_pos_a_FE}(d)-(f) we show the corresponding free energy difference for each order parameter solution (in case there are more than one). All plots demonstrate that there is only one physically relevant solution (with the biggest value of $\Delta_0$), for which $\delta\Omega < 0$. Moreover if the impurity density is far enough from the point where $\delta\Omega = 0$ [thin vertical line in Fig. \ref{Delta_vs_Gamma_aligned_pos_a_FE}(b)], the order parameter goes gradually to zero indicating that the phase transition is of second order, see Figs. \ref{Delta_vs_T_aligned_pos_a_FE}(a)-(b). Otherwise, the order parameter goes abruptly to zero at the critical point where the free energy difference becomes positive and the phase transition is first-order, as in Fig \ref{Delta_vs_T_aligned_pos_a_FE}(c). Changing the type of local exchange scattering off impurities by inverting the sign of $\alpha$ does not alter qualitatively the results discussed above and it is not shown here.

\subsection{Density of states}

\subsubsection{Randomly oriented impurities}
\begin{figure}[t]
\includegraphics[width=\columnwidth]{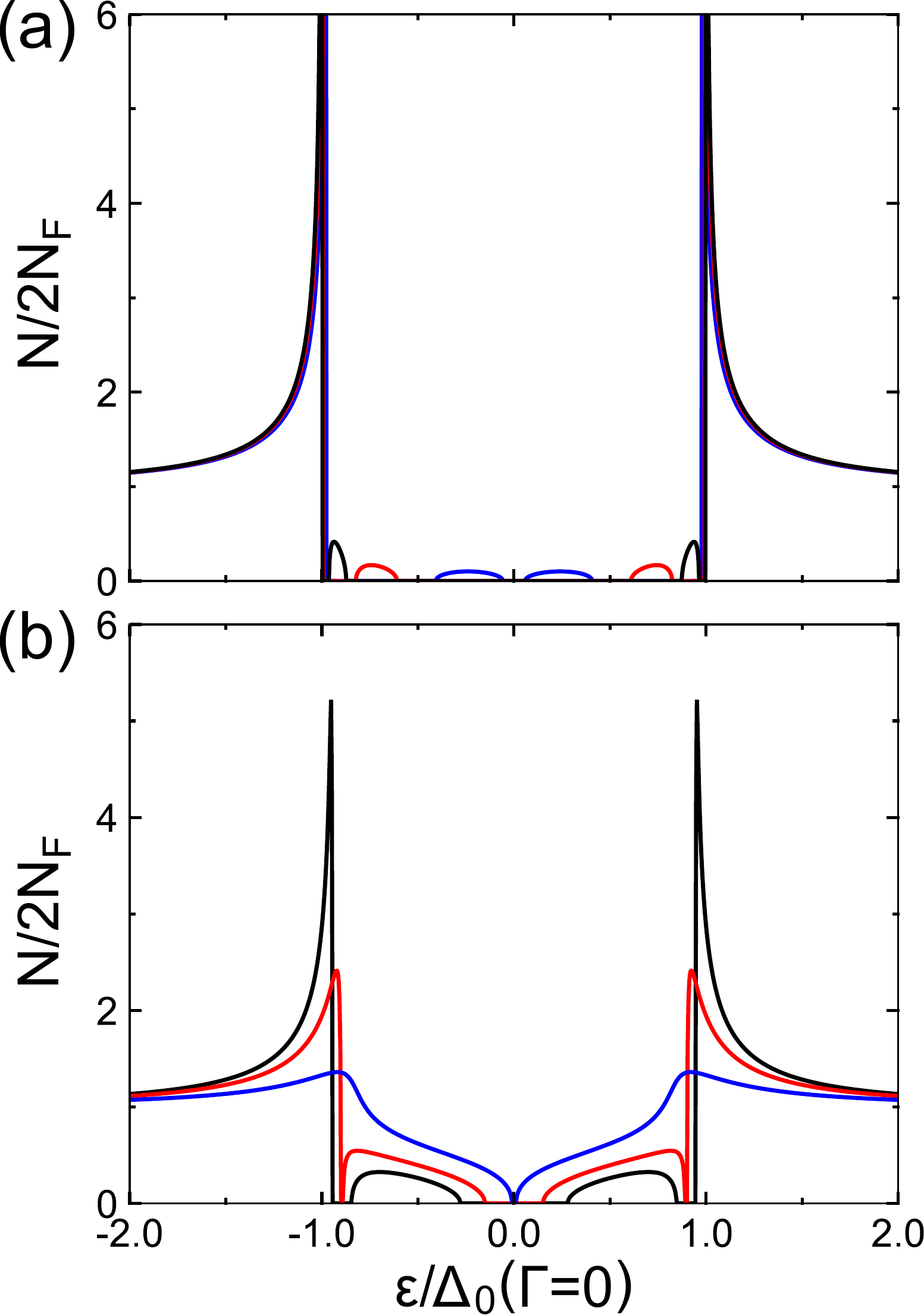}
\caption{(Color online) Density of states for randomly oriented impurity spins. (a): Evolution with $u_{\mathrm{S}}$. Black, red and blue lines correspond to $u_{\mathrm{S}}=2,4,8$ and $\Gamma/2\pi k_B T_{c0}=0.01$. (b) Evolution with $\Gamma$. Black, red and blue lines correspond to $\Gamma/2\pi k_B T_{c0}=0.05,0.1,0.2$ and $u_{\mathrm{S}}=5$. For both plots temperature is $T=0.01T_{c0}$.}
	\label{DOS_random}
\end{figure}

For the case of unpolarized impurities opposite-spin densities of states are the same and we therefore plot the total density of states as a function of energy in Fig. \ref{DOS_random}. Figure \ref{DOS_random}(a) demonstrates evolution of the density of states with varying impurity potential strength $u_{\mathrm{S}}$, while Fig. \ref{DOS_random}(b) shows its evolution with impurity density $\Gamma$. As one can see, the impurity potential strength sets the position of the YSR bands inside the gap, keeping the total number of YSR states fixed. On the other hand the density of impurities sets the size of the bands, by increasing the overall number of YSR states. Figure \ref{DOS_random}(b) also shows that randomly oriented impurity spins play a pair-breaking role in the system, decreasing the electron-hole coherence. The latter result was also obtained by Abrikosov and Gor'kov \cite{Abrikosov_Gorkov_1961} in the Born limit. They demonstrated that the order parameter and the energy gap in the spectrum are, generally speaking, two different quantities and there is a range of impurity densities where the so-called gapless superconductivity emerges \cite{Abrikosov_Gorkov_1961}.

\subsubsection{Ferromagnetically ordered impurities}
\begin{figure*}[t]
\includegraphics[width=\textwidth]{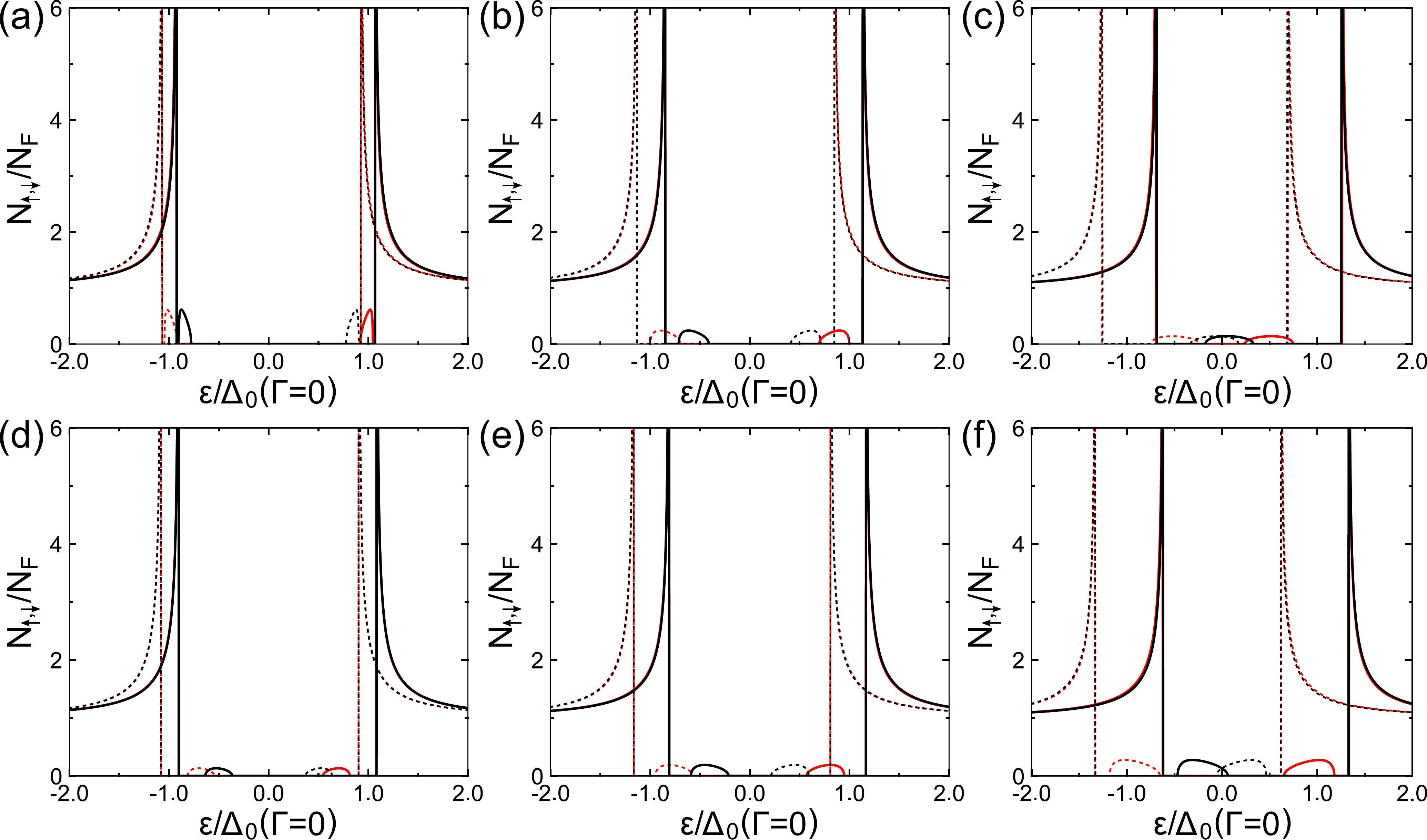}
\caption{(Color online) Spin-up (full line) and spin-down (dotted line) densities of states for ferromagnetically aligned impurity spins. Black lines correspond to $\alpha>0$ and red ones to $\alpha<0$. (a)-(c) show the effect of increasing $u_{\mathrm{S}}=2,4,8$ with $\Gamma/2\pi k_B T_{c0}=0.01$. (d)-(f) show the effect of increasing $\Gamma/2\pi k_B T_{c0}=0.005,0.01,0.02$ with $u_S=5$. Temperature is $T=0.01T_{c0}$.}
	\label{DOS_aligned}
\end{figure*}

For the case of ferromagnetically ordered impurities, opposite-spin densities of states are distinct and we plot them separately in Fig \ref{DOS_aligned}. The most important difference from the previous case is that YSR impurity bands are now spin-polarized, which allows for a non-zero magnetization in the system, as we will see below. Another complication arises from the background homogeneous exchange field generated by the impurities [see Eq. (\ref{Himp_ferro})], which shifts the opposite-spin spectra with respect to each other. The latter circumstance is important for understanding the weakening of superconductivity in this case. Indeed, if one increases impurity strength $u_{\mathrm{S}}$, the energy shift due to the exchange field drives the system to the normal state similarly to the effect of a pure Zeeman interaction \cite{Clogston1962,Chandrasekhar1962}. On the other hand, increasing the density of impurities results in a simultaneous growth of the impurity band and a shift due to the exchange field, which leads to faster decrease of superconducting properties than in the unpolarized case. Comparing Fig. \ref{Delta_vs_Gamma_and_T_random}(a) and Fig. \ref{Delta_vs_Gamma_aligned_pos_a_FE}, one can see that allowed densities of magnetic impurities for this case are one order of magnitude smaller than those for the unpolarized case.

Next, we compare how the type of local exchange scattering off impurities (see Fig. \ref{Fig2}) changes the spectral properties of the system. For the case of anti-ferromagnetic scattering (black lines in Fig. \ref{DOS_aligned}), $\alpha>0$, the spin-up YSR impurity subband splits off from the quasiparicle continuum at negative energies (solid lines), while spin-down sets in at positive energies (dotted lines). At the same time for ferromagnetic interaction (red lines in Fig. \ref{DOS_aligned}), $\alpha<0$, the opposite-spin YSR subbands exchange places. As can be seen from Figs. \ref{DOS_aligned}(a)-(c), the Zeeman-like shift due to the background exchange field counteracts the impurity band shift imposed by varying $u_{\mathrm{S}}$ for $\alpha<0$, while the two effects cooperate for $\alpha>0$. In Figs. \ref{DOS_aligned}(d)-(f) we observe that the background exchange field shifts the growing impurity subbands in the opposite directions for $\alpha>0$ and $\alpha<0$. All these feature are responsible for the differences in the transition temperature $T_c$ in Fig. \ref{Tc_vs_Gamma_aligned_neg_a}.
%
%\begin{figure*}[!t]
%\includegraphics[width=2\columnwidth]{plot_DOS_us_Gamma_combined_ordered_neg_aV1}
%\caption{The spin up (full line) and spin down (dotted line) density of states for aligned impurity spins and antiferromagnetic coupling. (a)-(c) show the effect of increasing $\Gamma/2\pi k_B T_{c0}=0.005,0.010,0.020$ with $u_S=5$. (d)-(f) show the effect of increasing $u_S=2,4,8$ with $\Gamma/2\pi k_B T_{c0}=0.01$. The temperature is taken to be $T=0.01T_{c0}$.}
%	\label{DoS_aligned_neg_a}
%\end{figure*}
%
%\FloatBarrier
\subsection{Magnetization and Sakurai phase transition}
When spins of magnetic impurities are ferromagnetically aligned, the single-particle spectrum of the system is not spin-degenerate, as we have seen in the previous section. This manifests itself as an imbalance between occupations of opposite-spin subbands, resulting in net magnetization. Moreover, since the in-gap YSR impurity bands are spin-polarized (see Fig. \ref{DOS_aligned}), we demonstrate below how one can observe the signature of the quantum phase transition first discussed by Sakurai in Ref. [\onlinecite{Sakurai1970}]. In this seminal paper it was shown that in presence of a single YSR magnetic impurity inside a superconductor there is a quantum phase transition in the ground state of the system when the impurity-induced YSR state crosses the Fermi level. By increasing the effective coupling strength $\zeta=\pi N_FJS$, where $J$ is the exchange interaction constant and $S$ is the impurity spin, the YSR state moves from one side of single-particle continuum to the opposite one, crossing zero when $\zeta = 1$. The qualitative physical picture of the phase transition is as follows. Since the YSR state is spin-polarized, when it gets occupied a single itinerant electron from the superconductor is bound to the impurity site in a singlet or triplet state depending on the nature of the local exchange interaction, see Fig. \ref{Fig2}. At the same time its time-reversed mate is left with an uncompensated spin and the ground state of the system in this case always contains a single quasiparticle \cite{Salkola1997,Balatsky2006}.

In terms of the parameters of our model, we have $\pi N_FJS \equiv \alpha u_{\mathrm{S}}$ and, in addition, there is a finite density of impurities $n = \pi N_F\Gamma$, instead of a single impurity. The parameter $\alpha$ has a meaning of the tunneling amplitude onto the impurity site, which means that the parameters that can be controlled, in principle, are $u_{\mathrm{S}}$ and $\Gamma$. Using Eq. (\ref{M_eqn}) we compute the magnetization $M$. By definition, the total magnetic moment in the system is $\mathcal{M} = MA$, where $A$ is the volume of the system. On the other hand, the total magnetic moment is related to the total spin of the system (in units of $\hbar$) $\mathcal{S}$ via $\mathcal{M} = -g\mu_B\mathcal{S}$, where g is the quasiparticle g-factor and $\mu_B$ is the Bohr magneton. Then we can introduce the average spin per magnetic impurity $\bar{s} \equiv \mathcal{S}/N$ via
\begin{align}
M = -g\pi N_F\Gamma\mu_B\bar{s}. \label{s_bar}
\end{align}
\begin{figure}[t]
	\includegraphics[width=\columnwidth]{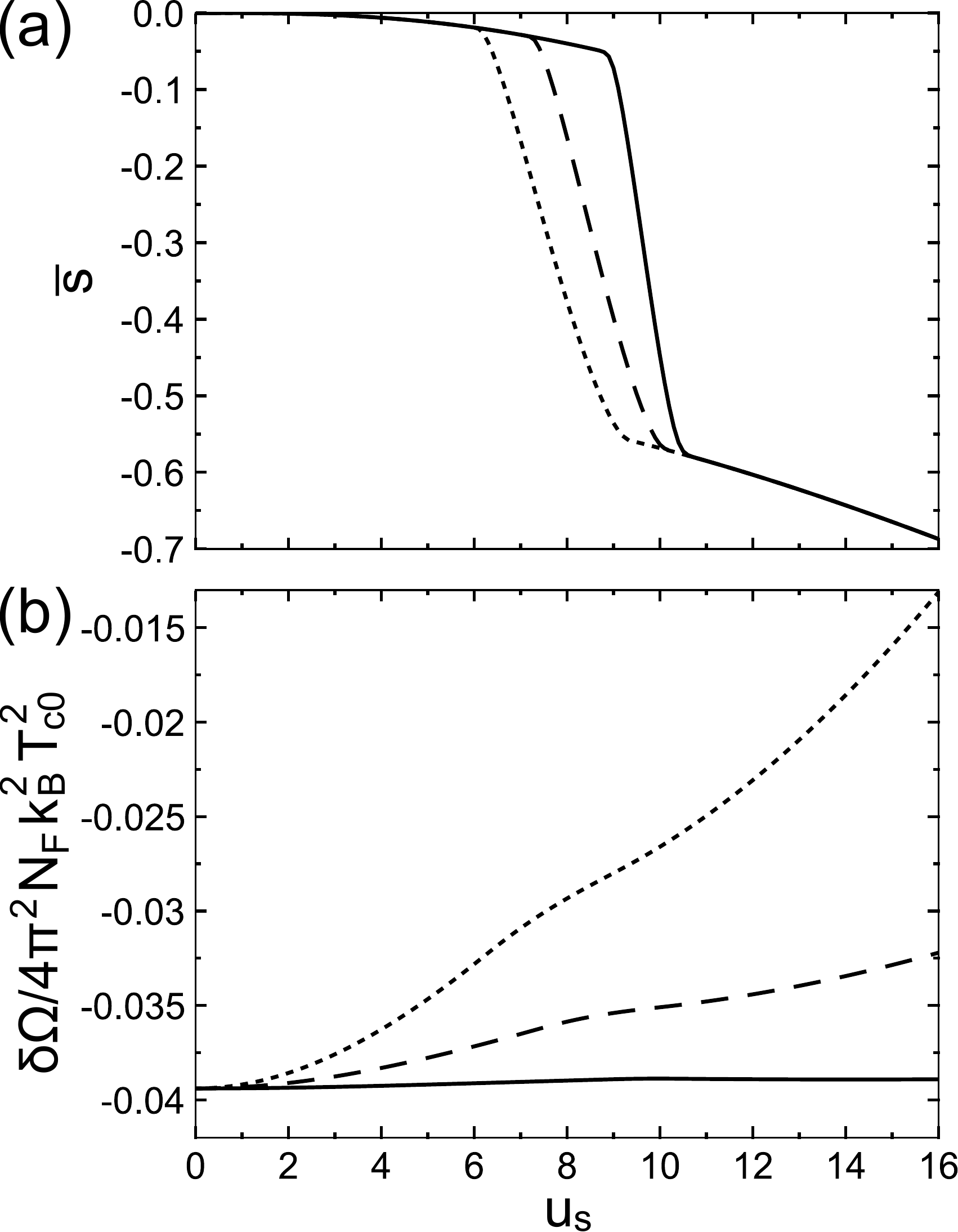}
	\caption{(a) Average spin per impurity $\bar{s}$ [see Eq. (\ref{s_bar})] as a function of $u_{\mathrm{S}}$ for $\alpha>0$. The impurity density $\Gamma/2\pi k_B T_{c0}$ is equal to $0.001$ (solid line), $0.005$ (dashed line) and $0.01$ (dotted line). (b) The corresponding Gibbs free energy difference.}
\label{M_vs_us_pos_a}
\end{figure}

In Fig. \ref{M_vs_us_pos_a}(a) we plot $\bar{s}$ as a function of impurity strength $u_{\mathrm{S}}$. As we know from the discussion of the density of states above, varying $u_{\mathrm{S}}$ changes the position of the YSR impurity band inside the gap, see Figs. \ref{DOS_aligned}(a)-(c). Therefore by tuning this parameter it is possible to make the impurity band cross the Fermi energy, which would mean that each magnetic impurity binds a single electron to itself, according to the qualitative picture of the phase transition discussed above. Then the average spin per impurity which is left uncompensated after the phase transition should be $\bar{s} = \mp 1/2$ for $\mathrm{sgn}(\alpha) = \pm 1$. As can be seen from Fig. \ref{M_vs_us_pos_a}(a), there is indeed a jump in $\bar{s}$ exactly equal to this value, but with a finite slope determined by the impurity density $\Gamma$. In addition, there is a background magnetization which makes the magnitude of $\bar{s}$ slowly increasing before and after the transition. It is determined by the difference in the densities of states in the (negative-energy) quasiparticle continuum, while the jump is due to the YSR impurity band. This difference is hardly seen in Fig. \ref{DOS_aligned}, but the more the YSR band splits off, the more it deforms the continuum it originated in, and the bigger the difference between the opposite-spin densities of states becomes. This additional feature was left out in the qualitative picture proposed by Sakurai \cite{Sakurai1970}. The free energy difference in Fig. \ref{M_vs_us_pos_a}(b) shows that the system remains superconducting after the phase transition. If $\alpha<0$, the results look similar, except that $\bar{s}\geq 0$.

\begin{figure}[t]
	\includegraphics[width=\columnwidth]{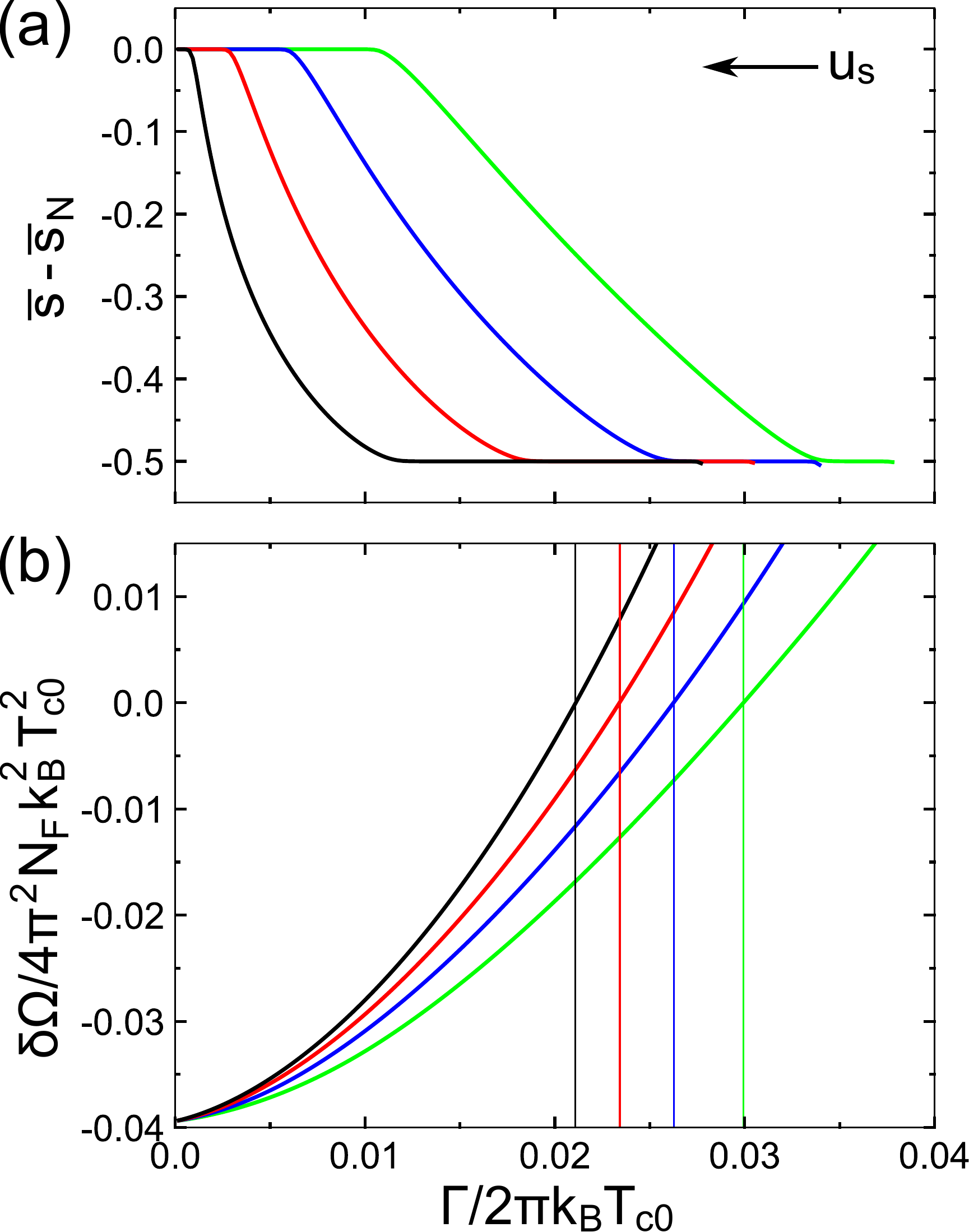}
\caption{(Color online) (a) Difference in the average spin (per impurity) in the superconducting $\bar{s}$ and normal $\bar{s}_N$ phases as a function of impurity density for $\alpha>0$. The impurity strength, from right to left, is $u_{\mathrm{S}}=6,7,8,9$. (b) The corresponding Gibbs free energy difference. Thin vertical lines depict the critical density at which it crosses zero.}
\label{M_vs_Gamma_pos_a}
\end{figure}

In Fig. \ref{M_vs_Gamma_pos_a} we perform a similar analysis, but we vary impurity density instead of their strength. This case is more suitable for experiments since controlling the density is easier. Moreover, in this case the phase transition is much more clearly observed. Indeed, in Fig. \ref{M_vs_Gamma_pos_a}(a) we plot the difference between the average spin per impurity in superconducting $\bar{s}$ and normal $\bar{s}_{\mathrm{N}}$ states. The latter is given by the first term on the rhs of Eq. (\ref{M_eqn}). One can see a clear jump in the figure with amplitude equal to $-1/2$, as expected. We already know that varying $\Gamma$ makes the YSR impurity band grow and shift (due to the background exchange field) at the same time, see Figs. \ref{DOS_aligned}(d)-(f). Thus, there is a limited range of parameters where one can see both plateaus, as in Fig. \ref{M_vs_Gamma_pos_a}(a). Moreover, because of the pair-breaking effect of the impurities, some of the results are not observable because of transition to the normal state, indicated by the free energy difference in Fig. \ref{M_vs_Gamma_pos_a}(b). Finally, if $\alpha<0$, in order to see the jump associated to the phase transition by varying $\Gamma$, one has to start in a state when the transition has occured (by choosing high enough $u_{\mathrm{S}}$, see Fig. \ref{DOS_aligned}). Then, gradually increasing $\Gamma$ the impurity band gets empty instead of becoming occupied (as for $\alpha>0$), and the initial plateau at $1/2$ evolves to the final one at zero.

%
%\begin{figure}[t]
%	\includegraphics[width=0.9\columnwidth]{plot_M_fn_Gamma_FE_ordered_neg_aV2}
%\caption{(a) The difference in the average number of spins per impurity between the superconducting and normal phase as a function of the impurity density for the case of antiferromagnetic coupling. The impurity potential is from left to right $u_S=13,14,15,16$. (b) The corresponding Gibbs free energy difference. The vertical line marks where it crosses zero.}
%\label{M_vs_Gamma_neg_a}
%\end{figure}
%
%\FloatBarrier

\section{Discussion and conclusions}\label{Sec_Concl}

Let us now consider in more detail our theoretical model and its range of validity. We start from a discussion of the necessity to take into account the background magnetic field collectively created by the distribution of magnetic impurities, see Eq. (\ref{Himp_ferro}). This exchange field makes the interaction of itinerant electrons with the magnetic impurities non-local. This is expected to hold if the impurity magnetic moment increases, since the magnetic field created by a magnetic dipole decays as $1/r^3$ as a function of distance $r$ from the dipole center. In our model increase of the magnetic moment of impurities corresponds to increasing $u_{\mathrm{S}}$. An interesting feature of the t-matrix approximation, see Appendix \ref{App_t_matr}, is that the {\em single-impurity} t-matrix $\hat{t}_{\mathrm{imp}}(\epsilon_n)$ becomes spin-independent if $u_{\mathrm{S}}\rightarrow\infty$ (unitary scattering limit), see Eq.(\ref{SE_eqn_ord}). This is a formal result and it means that in this limit, the magnetic impurities would behave just like scalar scatterers, which are known to satisfy the Anderson's theorem \cite{Anderson1956}. This would lead to a paradox that for the case of ferromagnetically ordered impurities, if we increase their strength $u_{\mathrm{S}}$, we recover the clean-limit results for the order parameter and other properties of the system \footnote{Note that the same result also holds for unpolarized magnetic impurities. We think that even if there is no net magnetic field in this case, the local exchange field created by each individual impurity would destroy superconductivity before the unitary limit is reached.}. The paradox is solved when we include the magnetic field generated by the ordered impurity moments, see Eq.(\ref{Himp_ferro}). In this case the order parameter gets suppressed long before we achieve the unitary limit. Note that this complication is absent when impurities are treated in the Born limit \cite{Gorkov_Rusinov_1964,Fulde_Maki_1966,Izyumov1974} because then the only free parameter is the impurity density (the impurity strength is assumed small).

We would also like to comment on the description of the impurity subsystem. In this work we considered the two limiting cases: (i) unpolarized and (ii) ferromagnetically ordered impurities. We did not include any theoretical description of how the ferromagnetic ordering takes place. However, as soon as the transition to the polarized case has occured, our results should be valid. A rigorous way of solving this problem would require to include, for example, a Heisenberg model for magnetic impurities and describe the dynamics of the coupled superconductor-impurities system. This would allow to consider a transient regime when the impurities are partially polarized. At the same time one would be able to introduce a new energy scale $k_BT_{\mathrm{Curie}}$, related to the Curie temperature of the impurity subsystem \cite{Gorkov_Rusinov_1964,Izyumov1974}. Then, our results for ferromagnetically aligned impurity spins discussed in this work are valid for $T\ll T_{\mathrm{Curie}}$.
Another point related to the description of impurities which was omitted here is the possibility to form clusters. To our knowledge, this effect cannot be automatically included into the quasiclassical theory described in this paper, and it would require a separate treatment. Allowing the magnetic impurities to cluster would smear out the sharp edges of the YSR impurity bands via the so-called Lifshitz tails \cite{Balatsky1997}.

Before conclusion we would like to comment on the order of magnitude of the impurity magnetic moment used in this work, corresponding to the value of $u_{\mathrm{S}}$. Assuming a quadratic energy dispersion of electrons in the normal state and taking their effective mass equal to the bare electron mass, we can estimate that for a superconductor with Fermi energy $E_F\simeq 10$~eV, the typical size of an impurity is $\sim 1-8$~nm, and its magnetic moment $\mathcal{M}\sim 8.5\times10^4~\mu_B$ (for $u_{\mathrm{S}} = 1$). This coincides with the typical size of small ferromagnetic islands used in modern experiments \cite{Asulin2009,Kim2014}. On the other hand it justifies our treatment of the impurity spins as classical \footnote{Surprisingly, even when the magnetic impurity is just a single atom, the YSR model can still be successfully used to analyze experimental data, see Ref. [\onlinecite{Ruby_2015}], even though a more complicated physics related to Kondo effect can emerge \cite{Liu1965,Griffin1965,Maki1967,Soda1967,Fowler1967,Abrikosov1969,Takano1969,Kitamura1970,Fowler1970,MullerHartmann1971,Matsuura1977,Ichinose1977,Ichinose1977_2,Franke2011}.}.

In conclusion, we have studied thermodynamic properties of a superconductor with a finite density of magnetic impurities, described within a generalized Yu-Shiba-Rusinov model (self-consistent $t$-matrix approximation). When the impurity spins are randomly oriented, most of the results are similar to the ones obtained by Abrikosov and Gor'kov \cite{Abrikosov_Gorkov_1961} within the first-order Born approximation. The only difference is that the YSR impurity bands are split off from the quasiparticle continuum. For the case of ferromagnetically ordered impurities we argue that it is necessary to include a background magnetic exchange field created by their spins. We have found that in this case the superconducting transition changes from second order to first order as the impurity strength is increased. At a critical impurity strength, superconductivity disappears. We have shown that the signature of the quantum phase transition of the system ground state due to the YSR impurity band crossing the Fermi energy is a drop in the magnetization. We emphasize that the initial idea of this phenomenon was put forward by Sakurai \cite{Sakurai1970} for a single impurity, while we have shown how it manifests itself for a finite impurity density.

\begin{acknowledgments}
The authors would like to acknowledge financial support from the Swedish Research Council. D.~P. and O.~S. contributed equally to this work.
\end{acknowledgments}

\appendix

\section{Impurity self-energy in t-matrix approximation}\label{App_t_matr}
In terms of the Matsubara Green's function, the impurity self-energy in t-matrix approximation is given by \cite{Serene_Rainer_PhysRep1983}
\begin{align}
\hat{h}_{\mathrm{t-matr}}(\epsilon_n,\mathbf{p}_{F}) = n\hat{t}_{\mathrm{imp}}(\epsilon_n,\mathbf{p}_{F},\mathbf{p}_{F}),
\end{align}
where the single-impurity t-matrix $\hat{t}_{\mathrm{imp}}$ satisfies
\begin{align}
&\hat{t}_{\mathrm{imp}}(\epsilon_n,\mathbf{p}_{F},\mathbf{p}_{F}^{\prime}) = \hat{v}(\mathbf{p}_{F},\mathbf{p}_{F}^{\prime})\label{t_matr_eqn}\\
&+N_F\int\frac{d\Omega_{\mathbf{p}_{F}^{\prime\prime}}}{4\pi}\hat{v}(\mathbf{p}_{F},\mathbf{p}_{F}^{\prime\prime})\hat{g}(\epsilon_n,\mathbf{p}_{F}^{\prime\prime})\hat{t}_{\mathrm{imp}}(\epsilon_n,\mathbf{p}_{F}^{\prime\prime},\mathbf{p}_{F}^{\prime}).\notag
\end{align}
\begin{center}
\begin{figure}[b]
\includegraphics[width=0.45\textwidth]{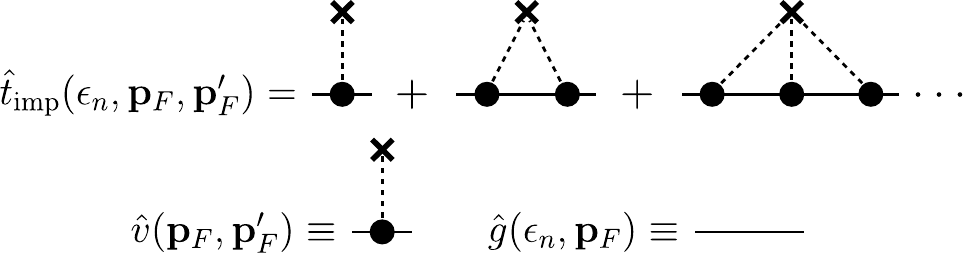}
\caption{Diagrammatic representation of equation (\ref{t_matr_eqn}).} \label{Fig_t_matr}
\end{figure}
\end{center}
Here $\hat{v}(\mathbf{p}_{F},\mathbf{p}_{F}^{\prime\prime})$ is the matrix element of the impurity potential between the quasiparticle states with momenta $\mathbf{p}_{F}$ and $\mathbf{p}_{F}^{\prime}$ on the Fermi surface (computed in the normal state of the system). Equation (\ref{t_matr_eqn}) is usually represented diagrammatically as a sum of diagrams for arbitrary number of quasiparticle scatterings on a given impurity, see Fig. \ref{Fig_t_matr}.
In this paper we consider only s-wave scattering off impurities, i.e. $\hat{v}(\mathbf{p}_{F},\mathbf{p}_{F}^{\prime\prime})$ is independent of momenta. For the two models of magnetic impurities described in the main text, see Eqs. (\ref{Himp_rand})-(\ref{Himp_ferro}), the matrix element $\hat{v}$ is
\begin{align}
\hat{v} = 
\begin{pmatrix}
v & 0\\
0 & v^{\ast}
\end{pmatrix},
\;\;v = v_0+\alpha v_{\mathrm{S}}\mathbf{m}\cdot\bs{\sigma}.
\end{align}
\subsection{Self-energy for randomly oriented impurities}
It is worth mentioning that for deriving Eq. (\ref{t_matr_eqn}) one has to perform averaging of the Dyson equation for the (full microscopic) propagator over the impurity positions defined as [see Eq. (\ref{Himp_rand})]
\begin{align}
\langle\bullet\rangle_{\mathrm{imp.\:pos.}} = \prod_{j=1}^{N}\int_{A}\frac{d\mathbf{r}_j}{A}\bullet,
\end{align}
where integration is performed over the system volume $A$. For the case of unpolarized magnetic impurities, besides averaging over impurity positions, one also has to average Eq. (\ref{t_matr_eqn}) over the magnetic moment directions. This is performed by parameterizing the unit vector $\mathbf{m}$ in spherical coordinates and defining
\begin{align}
\langle\bullet\rangle_{\mathrm{spin\:dir.}} = \int\frac{d\Omega_{\mathbf{m}}}{4\pi}\bullet.
\end{align}
Thus, using equations (\ref{GRiccati}), (\ref{gamma_structure_rand}) and (\ref{t_matr_eqn}) one can obtain the self-energy 
\begin{align}
\hat{h}_{\mathrm{imp}}(\epsilon_n) = n\langle\hat{t}_{\mathrm{imp}}(\epsilon_n)\rangle_{\mathrm{spin\:dir.}},
\end{align}
with matrix elements written in Eq. (\ref{SE_imps_rand}).
\subsection{Self-energy for ferromagnetically ordered impurities}
For the case of ferromagnetically ordered magnetic impurities we can choose the coordinate system in spin space such as $\mathbf{m}_j \equiv \mathbf{m} = (0,0,1)$. Since in this case, apart from the local scattering by the impurities, we also have a background magnetic field in the system [see Eq. (\ref{Himp_ferro})], the impurity self-energy consists of two parts
\begin{align}
\hat{h}_{\mathrm{imp}}(\epsilon_n) = \beta n v_{\mathrm{S}}\sigma_{3}\hat{1}+n\hat{t}_{\mathrm{imp}}(\epsilon_n),
\end{align}
where the first term has a form of Zeeman interaction, while the second one is obtained by solving Eq. (\ref{t_matr_eqn}).

\section{Free energy functional}\label{App_FE}
In this section we briefly describe how to compute the difference between the Gibbs free energies in superconducting and normal states. The free energy is a functional of the quasiclassical propagator and self-energies. So, we define
\begin{align}
\delta\Omega[\hat{g},\hat{h},T] = \Omega_S[\hat{g},\hat{h},T] - \Omega_N[\hat{g},\hat{h},T],
\end{align} 
which means that for $\delta\Omega < 0$, the superconducting state is more energetically favorable. In order to derive Eq. (\ref{FE_eqn}) we follow Refs. \cite{Serene_Rainer_PhysRep1983,Keller1988,Sauls_FLT1994} and write down
\begin{align}
&\Omega_S[\hat{g},\hat{h},T] = \Omega_N[\hat{g},\hat{h},T = 0]\notag\\
&-\frac{1}{2}N_f\int\frac{d\Omega_{\mathbf{p}_{F}}}{4\pi}k_BT\sum_{|\epsilon_n|<\epsilon_c}\!\mathrm{Tr}
\left\{\vphantom{\int\limits_{-\epsilon_c}^{\epsilon_c}}\hat{h}(\mathbf{p}_{F},\epsilon_n)\hat{g}(\mathbf{k},\epsilon_n)\right.\notag\\
&\left.+\int\limits_{-\epsilon_c}^{\epsilon_c}\!\!d\xi_{\mathbf{k}}\ln\!\left[-\hat{G}_0^{-1}(\mathbf{k},\epsilon_n)+\hat{h}(\mathbf{p}_{F},\epsilon_n)\right]\right\}+\delta\Phi\left[\hat{g}\right].
\label{Omega_S}
\end{align}
Here $\hat{G}_0^{-1}(\mathbf{k},\epsilon_n) = i\epsilon_n\hat{\tau}_3-\xi_{\mathbf{k}}$, where $\xi_{\mathbf{k}}$ is a single-particle spectrum in the normal state (calculated with respect to the Fermi energy $E_F$). $\delta\Phi\left[\hat{g}\right] = \Phi_S\left[\hat{g}\right]-\Phi_N\left[\hat{g}\right]$, where $\Phi\left[\hat{g}\right]$ is a functional which generates the perturbation expansion for the skeleton self-energy diagrams \cite{Serene_Rainer_PhysRep1983,Sauls_FLT1994}. We note that the log-term on the rhs of Eq. (\ref{Omega_S}) contains a finite temperature contribution to the normal state free energy $\Omega_N[\hat{g},\hat{h},T > 0]$, which has to be subtracted when computing the integral \cite{Serene_Rainer_PhysRep1983}. Equation (\ref{FE_eqn}) is obtained from Eq. (\ref{Omega_S}) using the quasiclassical self-energy $\hat{h}$ appropriate to our model and assuming quadratic energy spectrum in the normal state $\xi_{\mathbf{k}} = \hbar^2k^2/2m^{\ast}-E_F$, where $m^{\ast}$ is the effective mass.

\bibliography{paper_YSR_bibliography}

\begin{thebibliography}{80}%
\makeatletter
\providecommand \@ifxundefined [1]{%
 \@ifx{#1\undefined}
}%
\providecommand \@ifnum [1]{%
 \ifnum #1\expandafter \@firstoftwo
 \else \expandafter \@secondoftwo
 \fi
}%
\providecommand \@ifx [1]{%
 \ifx #1\expandafter \@firstoftwo
 \else \expandafter \@secondoftwo
 \fi
}%
\providecommand \natexlab [1]{#1}%
\providecommand \enquote  [1]{``#1''}%
\providecommand \bibnamefont  [1]{#1}%
\providecommand \bibfnamefont [1]{#1}%
\providecommand \citenamefont [1]{#1}%
\providecommand \href@noop [0]{\@secondoftwo}%
\providecommand \href [0]{\begingroup \@sanitize@url \@href}%
\providecommand \@href[1]{\@@startlink{#1}\@@href}%
\providecommand \@@href[1]{\endgroup#1\@@endlink}%
\providecommand \@sanitize@url [0]{\catcode `\\12\catcode `\$12\catcode
  `\&12\catcode `\#12\catcode `\^12\catcode `\_12\catcode `\%12\relax}%
\providecommand \@@startlink[1]{}%
\providecommand \@@endlink[0]{}%
\providecommand \url  [0]{\begingroup\@sanitize@url \@url }%
\providecommand \@url [1]{\endgroup\@href {#1}{\urlprefix }}%
\providecommand \urlprefix  [0]{URL }%
\providecommand \Eprint [0]{\href }%
\providecommand \doibase [0]{http://dx.doi.org/}%
\providecommand \selectlanguage [0]{\@gobble}%
\providecommand \bibinfo  [0]{\@secondoftwo}%
\providecommand \bibfield  [0]{\@secondoftwo}%
\providecommand \translation [1]{[#1]}%
\providecommand \BibitemOpen [0]{}%
\providecommand \bibitemStop [0]{}%
\providecommand \bibitemNoStop [0]{.\EOS\space}%
\providecommand \EOS [0]{\spacefactor3000\relax}%
\providecommand \BibitemShut  [1]{\csname bibitem#1\endcsname}%
\let\auto@bib@innerbib\@empty
%</preamble>
\bibitem [{\citenamefont {Felner}\ \emph {et~al.}(1997)\citenamefont {Felner},
  \citenamefont {Asaf}, \citenamefont {Levi},\ and\ \citenamefont
  {Millo}}]{Felner1997}%
  \BibitemOpen
  \bibfield  {author} {\bibinfo {author} {\bibfnamefont {I.}~\bibnamefont
  {Felner}}, \bibinfo {author} {\bibfnamefont {U.}~\bibnamefont {Asaf}},
  \bibinfo {author} {\bibfnamefont {Y.}~\bibnamefont {Levi}}, \ and\ \bibinfo
  {author} {\bibfnamefont {O.}~\bibnamefont {Millo}},\ }\href {\doibase
  10.1103/PhysRevB.55.R3374} {\bibfield  {journal} {\bibinfo  {journal} {Phys.
  Rev. B}\ }\textbf {\bibinfo {volume} {55}},\ \bibinfo {pages} {R3374}
  (\bibinfo {year} {1997})}\BibitemShut {NoStop}%
\bibitem [{\citenamefont {Saxena}\ \emph {et~al.}(2000)\citenamefont {Saxena},
  \citenamefont {Agarwal}, \citenamefont {Ahilan}, \citenamefont {Grosche},
  \citenamefont {Haselwimmer}, \citenamefont {Steiner}, \citenamefont {Pugh},
  \citenamefont {Walker}, \citenamefont {Julian}, \citenamefont {Monthoux},
  \citenamefont {Lonzarich}, \citenamefont {Huxley}, \citenamefont {Sheikin},
  \citenamefont {Braithwaite},\ and\ \citenamefont {Flouquet}}]{Saxena2000}%
  \BibitemOpen
  \bibfield  {author} {\bibinfo {author} {\bibfnamefont {S.~S.}\ \bibnamefont
  {Saxena}}, \bibinfo {author} {\bibfnamefont {P.}~\bibnamefont {Agarwal}},
  \bibinfo {author} {\bibfnamefont {K.}~\bibnamefont {Ahilan}}, \bibinfo
  {author} {\bibfnamefont {F.~M.}\ \bibnamefont {Grosche}}, \bibinfo {author}
  {\bibfnamefont {R.~K.~W.}\ \bibnamefont {Haselwimmer}}, \bibinfo {author}
  {\bibfnamefont {M.~J.}\ \bibnamefont {Steiner}}, \bibinfo {author}
  {\bibfnamefont {E.}~\bibnamefont {Pugh}}, \bibinfo {author} {\bibfnamefont
  {I.~R.}\ \bibnamefont {Walker}}, \bibinfo {author} {\bibfnamefont {S.~R.}\
  \bibnamefont {Julian}}, \bibinfo {author} {\bibfnamefont {P.}~\bibnamefont
  {Monthoux}}, \bibinfo {author} {\bibfnamefont {G.~G.}\ \bibnamefont
  {Lonzarich}}, \bibinfo {author} {\bibfnamefont {A.}~\bibnamefont {Huxley}},
  \bibinfo {author} {\bibfnamefont {I.}~\bibnamefont {Sheikin}}, \bibinfo
  {author} {\bibfnamefont {D.}~\bibnamefont {Braithwaite}}, \ and\ \bibinfo
  {author} {\bibfnamefont {J.}~\bibnamefont {Flouquet}},\ }\href {\doibase
  10.1038/35020500} {\bibfield  {journal} {\bibinfo  {journal} {Nature}\
  }\textbf {\bibinfo {volume} {406}},\ \bibinfo {pages} {587} (\bibinfo {year}
  {2000})}\BibitemShut {NoStop}%
\bibitem [{\citenamefont {Aoki}\ \emph {et~al.}(2001)\citenamefont {Aoki},
  \citenamefont {Huxley}, \citenamefont {Ressouche}, \citenamefont
  {Braithwaite}, \citenamefont {Flouquet}, \citenamefont {Brison},
  \citenamefont {Lhotel},\ and\ \citenamefont {Paulsen}}]{Aoki2001}%
  \BibitemOpen
  \bibfield  {author} {\bibinfo {author} {\bibfnamefont {D.}~\bibnamefont
  {Aoki}}, \bibinfo {author} {\bibfnamefont {A.}~\bibnamefont {Huxley}},
  \bibinfo {author} {\bibfnamefont {E.}~\bibnamefont {Ressouche}}, \bibinfo
  {author} {\bibfnamefont {D.}~\bibnamefont {Braithwaite}}, \bibinfo {author}
  {\bibfnamefont {J.}~\bibnamefont {Flouquet}}, \bibinfo {author}
  {\bibfnamefont {J.-P.}\ \bibnamefont {Brison}}, \bibinfo {author}
  {\bibfnamefont {E.}~\bibnamefont {Lhotel}}, \ and\ \bibinfo {author}
  {\bibfnamefont {C.}~\bibnamefont {Paulsen}},\ }\href {\doibase
  10.1038/35098048} {\bibfield  {journal} {\bibinfo  {journal} {Nature}\
  }\textbf {\bibinfo {volume} {413}},\ \bibinfo {pages} {613} (\bibinfo {year}
  {2001})}\BibitemShut {NoStop}%
\bibitem [{\citenamefont {Pfleiderer}\ \emph {et~al.}(2001)\citenamefont
  {Pfleiderer}, \citenamefont {Uhlarz}, \citenamefont {Hayden}, \citenamefont
  {Vollmer}, \citenamefont {Lohneysen}, \citenamefont {Bernhoeft},\ and\
  \citenamefont {Lonzarich}}]{Pfleiderer2001}%
  \BibitemOpen
  \bibfield  {author} {\bibinfo {author} {\bibfnamefont {C.}~\bibnamefont
  {Pfleiderer}}, \bibinfo {author} {\bibfnamefont {M.}~\bibnamefont {Uhlarz}},
  \bibinfo {author} {\bibfnamefont {S.~M.}\ \bibnamefont {Hayden}}, \bibinfo
  {author} {\bibfnamefont {R.}~\bibnamefont {Vollmer}}, \bibinfo {author}
  {\bibfnamefont {H.~v.}\ \bibnamefont {Lohneysen}}, \bibinfo {author}
  {\bibfnamefont {N.~R.}\ \bibnamefont {Bernhoeft}}, \ and\ \bibinfo {author}
  {\bibfnamefont {G.~G.}\ \bibnamefont {Lonzarich}},\ }\href {\doibase
  10.1038/35083531} {\bibfield  {journal} {\bibinfo  {journal} {Nature}\
  }\textbf {\bibinfo {volume} {412}},\ \bibinfo {pages} {58} (\bibinfo {year}
  {2001})}\BibitemShut {NoStop}%
\bibitem [{\citenamefont {Dikin}\ \emph {et~al.}(2011)\citenamefont {Dikin},
  \citenamefont {Mehta}, \citenamefont {Bark}, \citenamefont {Folkman},
  \citenamefont {Eom},\ and\ \citenamefont {Chandrasekhar}}]{Dikin2011}%
  \BibitemOpen
  \bibfield  {author} {\bibinfo {author} {\bibfnamefont {D.~A.}\ \bibnamefont
  {Dikin}}, \bibinfo {author} {\bibfnamefont {M.}~\bibnamefont {Mehta}},
  \bibinfo {author} {\bibfnamefont {C.~W.}\ \bibnamefont {Bark}}, \bibinfo
  {author} {\bibfnamefont {C.~M.}\ \bibnamefont {Folkman}}, \bibinfo {author}
  {\bibfnamefont {C.~B.}\ \bibnamefont {Eom}}, \ and\ \bibinfo {author}
  {\bibfnamefont {V.}~\bibnamefont {Chandrasekhar}},\ }\href {\doibase
  10.1103/PhysRevLett.107.056802} {\bibfield  {journal} {\bibinfo  {journal}
  {Phys. Rev. Lett.}\ }\textbf {\bibinfo {volume} {107}},\ \bibinfo {pages}
  {056802} (\bibinfo {year} {2011})}\BibitemShut {NoStop}%
\bibitem [{\citenamefont {Keizer}\ \emph {et~al.}(2006)\citenamefont {Keizer},
  \citenamefont {Goennenwein}, \citenamefont {Klapwijk}, \citenamefont {Miao},
  \citenamefont {Xiao},\ and\ \citenamefont {Gupta}}]{Keizer2006}%
  \BibitemOpen
  \bibfield  {author} {\bibinfo {author} {\bibfnamefont {R.~S.}\ \bibnamefont
  {Keizer}}, \bibinfo {author} {\bibfnamefont {S.~T.~B.}\ \bibnamefont
  {Goennenwein}}, \bibinfo {author} {\bibfnamefont {T.~M.}\ \bibnamefont
  {Klapwijk}}, \bibinfo {author} {\bibfnamefont {G.}~\bibnamefont {Miao}},
  \bibinfo {author} {\bibfnamefont {G.}~\bibnamefont {Xiao}}, \ and\ \bibinfo
  {author} {\bibfnamefont {A.}~\bibnamefont {Gupta}},\ }\href {\doibase
  10.1038/nature04499} {\bibfield  {journal} {\bibinfo  {journal} {Nature}\
  }\textbf {\bibinfo {volume} {439}},\ \bibinfo {pages} {825} (\bibinfo {year}
  {2006})}\BibitemShut {NoStop}%
\bibitem [{\citenamefont {Anwar}\ \emph {et~al.}(2010)\citenamefont {Anwar},
  \citenamefont {Czeschka}, \citenamefont {Hesselberth}, \citenamefont
  {Porcu},\ and\ \citenamefont {Aarts}}]{Anwar2010}%
  \BibitemOpen
  \bibfield  {author} {\bibinfo {author} {\bibfnamefont {M.~S.}\ \bibnamefont
  {Anwar}}, \bibinfo {author} {\bibfnamefont {F.}~\bibnamefont {Czeschka}},
  \bibinfo {author} {\bibfnamefont {M.}~\bibnamefont {Hesselberth}}, \bibinfo
  {author} {\bibfnamefont {M.}~\bibnamefont {Porcu}}, \ and\ \bibinfo {author}
  {\bibfnamefont {J.}~\bibnamefont {Aarts}},\ }\href {\doibase
  10.1103/PhysRevB.82.100501} {\bibfield  {journal} {\bibinfo  {journal} {Phys.
  Rev. B}\ }\textbf {\bibinfo {volume} {82}},\ \bibinfo {pages} {100501}
  (\bibinfo {year} {2010})}\BibitemShut {NoStop}%
\bibitem [{\citenamefont {Robinson}\ \emph {et~al.}(2010)\citenamefont
  {Robinson}, \citenamefont {Witt},\ and\ \citenamefont
  {Blamire}}]{Robinson2010}%
  \BibitemOpen
  \bibfield  {author} {\bibinfo {author} {\bibfnamefont {J.~W.~A.}\
  \bibnamefont {Robinson}}, \bibinfo {author} {\bibfnamefont {J.~D.~S.}\
  \bibnamefont {Witt}}, \ and\ \bibinfo {author} {\bibfnamefont {M.~G.}\
  \bibnamefont {Blamire}},\ }\href {\doibase 10.1126/science.1189246}
  {\bibfield  {journal} {\bibinfo  {journal} {Science}\ }\textbf {\bibinfo
  {volume} {329}},\ \bibinfo {pages} {59} (\bibinfo {year} {2010})}\BibitemShut
  {NoStop}%
\bibitem [{\citenamefont {Woolf}\ and\ \citenamefont {Reif}(1965)}]{Woolf1965}%
  \BibitemOpen
  \bibfield  {author} {\bibinfo {author} {\bibfnamefont {M.~A.}\ \bibnamefont
  {Woolf}}\ and\ \bibinfo {author} {\bibfnamefont {F.}~\bibnamefont {Reif}},\
  }\href {\doibase 10.1103/PhysRev.137.A557} {\bibfield  {journal} {\bibinfo
  {journal} {Phys. Rev.}\ }\textbf {\bibinfo {volume} {137}},\ \bibinfo {pages}
  {A557} (\bibinfo {year} {1965})}\BibitemShut {NoStop}%
\bibitem [{\citenamefont {Yazdani}\ \emph {et~al.}(1997)\citenamefont
  {Yazdani}, \citenamefont {Jones}, \citenamefont {Lutz}, \citenamefont
  {Crommie},\ and\ \citenamefont {Eigler}}]{Yazdani1997}%
  \BibitemOpen
  \bibfield  {author} {\bibinfo {author} {\bibfnamefont {A.}~\bibnamefont
  {Yazdani}}, \bibinfo {author} {\bibfnamefont {B.~A.}\ \bibnamefont {Jones}},
  \bibinfo {author} {\bibfnamefont {C.~P.}\ \bibnamefont {Lutz}}, \bibinfo
  {author} {\bibfnamefont {M.~F.}\ \bibnamefont {Crommie}}, \ and\ \bibinfo
  {author} {\bibfnamefont {D.~M.}\ \bibnamefont {Eigler}},\ }\href {\doibase
  10.1126/science.275.5307.1767} {\bibfield  {journal} {\bibinfo  {journal}
  {Science}\ }\textbf {\bibinfo {volume} {275}},\ \bibinfo {pages} {1767}
  (\bibinfo {year} {1997})}\BibitemShut {NoStop}%
\bibitem [{\citenamefont {Hudson}\ \emph {et~al.}(2001)\citenamefont {Hudson},
  \citenamefont {Lang}, \citenamefont {Madhavan}, \citenamefont {Pan},
  \citenamefont {Eisaki}, \citenamefont {Uchida},\ and\ \citenamefont
  {Davis}}]{Hudson2001}%
  \BibitemOpen
  \bibfield  {author} {\bibinfo {author} {\bibfnamefont {E.~W.}\ \bibnamefont
  {Hudson}}, \bibinfo {author} {\bibfnamefont {K.~M.}\ \bibnamefont {Lang}},
  \bibinfo {author} {\bibfnamefont {V.}~\bibnamefont {Madhavan}}, \bibinfo
  {author} {\bibfnamefont {S.~H.}\ \bibnamefont {Pan}}, \bibinfo {author}
  {\bibfnamefont {H.}~\bibnamefont {Eisaki}}, \bibinfo {author} {\bibfnamefont
  {S.}~\bibnamefont {Uchida}}, \ and\ \bibinfo {author} {\bibfnamefont {J.~C.}\
  \bibnamefont {Davis}},\ }\href {\doibase 10.1038/35082019} {\bibfield
  {journal} {\bibinfo  {journal} {Nature}\ }\textbf {\bibinfo {volume} {411}},\
  \bibinfo {pages} {920} (\bibinfo {year} {2001})}\BibitemShut {NoStop}%
\bibitem [{\citenamefont {Ji}\ \emph {et~al.}(2008)\citenamefont {Ji},
  \citenamefont {Zhang}, \citenamefont {Fu}, \citenamefont {Chen},
  \citenamefont {Ma}, \citenamefont {Li}, \citenamefont {Duan}, \citenamefont
  {Jia},\ and\ \citenamefont {Xue}}]{Ji2008}%
  \BibitemOpen
  \bibfield  {author} {\bibinfo {author} {\bibfnamefont {S.-H.}\ \bibnamefont
  {Ji}}, \bibinfo {author} {\bibfnamefont {T.}~\bibnamefont {Zhang}}, \bibinfo
  {author} {\bibfnamefont {Y.-S.}\ \bibnamefont {Fu}}, \bibinfo {author}
  {\bibfnamefont {X.}~\bibnamefont {Chen}}, \bibinfo {author} {\bibfnamefont
  {X.-C.}\ \bibnamefont {Ma}}, \bibinfo {author} {\bibfnamefont
  {J.}~\bibnamefont {Li}}, \bibinfo {author} {\bibfnamefont {W.-H.}\
  \bibnamefont {Duan}}, \bibinfo {author} {\bibfnamefont {J.-F.}\ \bibnamefont
  {Jia}}, \ and\ \bibinfo {author} {\bibfnamefont {Q.-K.}\ \bibnamefont
  {Xue}},\ }\href {\doibase 10.1103/PhysRevLett.100.226801} {\bibfield
  {journal} {\bibinfo  {journal} {Phys. Rev. Lett.}\ }\textbf {\bibinfo
  {volume} {100}},\ \bibinfo {pages} {226801} (\bibinfo {year}
  {2008})}\BibitemShut {NoStop}%
\bibitem [{\citenamefont {Asulin}\ \emph {et~al.}(2009)\citenamefont {Asulin},
  \citenamefont {Yuli}, \citenamefont {Koren},\ and\ \citenamefont
  {Millo}}]{Asulin2009}%
  \BibitemOpen
  \bibfield  {author} {\bibinfo {author} {\bibfnamefont {I.}~\bibnamefont
  {Asulin}}, \bibinfo {author} {\bibfnamefont {O.}~\bibnamefont {Yuli}},
  \bibinfo {author} {\bibfnamefont {G.}~\bibnamefont {Koren}}, \ and\ \bibinfo
  {author} {\bibfnamefont {O.}~\bibnamefont {Millo}},\ }\href {\doibase
  10.1103/PhysRevB.79.174524} {\bibfield  {journal} {\bibinfo  {journal} {Phys.
  Rev. B}\ }\textbf {\bibinfo {volume} {79}},\ \bibinfo {pages} {174524}
  (\bibinfo {year} {2009})}\BibitemShut {NoStop}%
\bibitem [{\citenamefont {Nadj-Perge}\ \emph {et~al.}(2013)\citenamefont
  {Nadj-Perge}, \citenamefont {Drozdov}, \citenamefont {Bernevig},\ and\
  \citenamefont {Yazdani}}]{Nadj-Perge2013}%
  \BibitemOpen
  \bibfield  {author} {\bibinfo {author} {\bibfnamefont {S.}~\bibnamefont
  {Nadj-Perge}}, \bibinfo {author} {\bibfnamefont {I.~K.}\ \bibnamefont
  {Drozdov}}, \bibinfo {author} {\bibfnamefont {B.~A.}\ \bibnamefont
  {Bernevig}}, \ and\ \bibinfo {author} {\bibfnamefont {A.}~\bibnamefont
  {Yazdani}},\ }\href {\doibase 10.1103/PhysRevB.88.020407} {\bibfield
  {journal} {\bibinfo  {journal} {Phys. Rev. B}\ }\textbf {\bibinfo {volume}
  {88}},\ \bibinfo {pages} {020407} (\bibinfo {year} {2013})}\BibitemShut
  {NoStop}%
\bibitem [{\citenamefont {Pientka}\ \emph {et~al.}(2013)\citenamefont
  {Pientka}, \citenamefont {Glazman},\ and\ \citenamefont {von
  Oppen}}]{Pientka2013}%
  \BibitemOpen
  \bibfield  {author} {\bibinfo {author} {\bibfnamefont {F.}~\bibnamefont
  {Pientka}}, \bibinfo {author} {\bibfnamefont {L.~I.}\ \bibnamefont
  {Glazman}}, \ and\ \bibinfo {author} {\bibfnamefont {F.}~\bibnamefont {von
  Oppen}},\ }\href {\doibase 10.1103/PhysRevB.88.155420} {\bibfield  {journal}
  {\bibinfo  {journal} {Phys. Rev. B}\ }\textbf {\bibinfo {volume} {88}},\
  \bibinfo {pages} {155420} (\bibinfo {year} {2013})}\BibitemShut {NoStop}%
\bibitem [{\citenamefont {Peng}\ \emph {et~al.}(2015)\citenamefont {Peng},
  \citenamefont {Pientka}, \citenamefont {Glazman},\ and\ \citenamefont {von
  Oppen}}]{Peng2015}%
  \BibitemOpen
  \bibfield  {author} {\bibinfo {author} {\bibfnamefont {Y.}~\bibnamefont
  {Peng}}, \bibinfo {author} {\bibfnamefont {F.}~\bibnamefont {Pientka}},
  \bibinfo {author} {\bibfnamefont {L.~I.}\ \bibnamefont {Glazman}}, \ and\
  \bibinfo {author} {\bibfnamefont {F.}~\bibnamefont {von Oppen}},\ }\href
  {\doibase 10.1103/PhysRevLett.114.106801} {\bibfield  {journal} {\bibinfo
  {journal} {Phys. Rev. Lett.}\ }\textbf {\bibinfo {volume} {114}},\ \bibinfo
  {pages} {106801} (\bibinfo {year} {2015})}\BibitemShut {NoStop}%
\bibitem [{\citenamefont {Nadj-Perge}\ \emph {et~al.}(2014)\citenamefont
  {Nadj-Perge}, \citenamefont {Drozdov}, \citenamefont {Li}, \citenamefont
  {Chen}, \citenamefont {Jeon}, \citenamefont {Seo}, \citenamefont {MacDonald},
  \citenamefont {Bernevig},\ and\ \citenamefont {Yazdani}}]{Nadj-Perge2014}%
  \BibitemOpen
  \bibfield  {author} {\bibinfo {author} {\bibfnamefont {S.}~\bibnamefont
  {Nadj-Perge}}, \bibinfo {author} {\bibfnamefont {I.~K.}\ \bibnamefont
  {Drozdov}}, \bibinfo {author} {\bibfnamefont {J.}~\bibnamefont {Li}},
  \bibinfo {author} {\bibfnamefont {H.}~\bibnamefont {Chen}}, \bibinfo {author}
  {\bibfnamefont {S.}~\bibnamefont {Jeon}}, \bibinfo {author} {\bibfnamefont
  {J.}~\bibnamefont {Seo}}, \bibinfo {author} {\bibfnamefont {A.~H.}\
  \bibnamefont {MacDonald}}, \bibinfo {author} {\bibfnamefont {B.~A.}\
  \bibnamefont {Bernevig}}, \ and\ \bibinfo {author} {\bibfnamefont
  {A.}~\bibnamefont {Yazdani}},\ }\href {\doibase 10.1126/science.1259327}
  {\bibfield  {journal} {\bibinfo  {journal} {Science}\ }\textbf {\bibinfo
  {volume} {346}},\ \bibinfo {pages} {602} (\bibinfo {year}
  {2014})}\BibitemShut {NoStop}%
\bibitem [{\citenamefont {Abrikosov}\ and\ \citenamefont
  {Gor'kov}(1960)}]{Abrikosov_Gorkov_1961}%
  \BibitemOpen
  \bibfield  {author} {\bibinfo {author} {\bibfnamefont {A.~A.}\ \bibnamefont
  {Abrikosov}}\ and\ \bibinfo {author} {\bibfnamefont {L.~P.}\ \bibnamefont
  {Gor'kov}},\ }\href@noop {} {\bibfield  {journal} {\bibinfo  {journal} {Zh.
  Eksp. Teor. Fiz.}\ }\textbf {\bibinfo {volume} {39}},\ \bibinfo {pages}
  {1781} (\bibinfo {year} {1960})},\ \bibinfo {note} {[Sov. Phys.--JETP {\bf
  12}, 1243 (1961)]}\BibitemShut {NoStop}%
\bibitem [{\citenamefont {Yu}(1965)}]{Yu1965}%
  \BibitemOpen
  \bibfield  {author} {\bibinfo {author} {\bibfnamefont {L.}~\bibnamefont
  {Yu}},\ }\href {\doibase 10.7498/aps.21.75} {\bibfield  {journal} {\bibinfo
  {journal} {Acta Phys. Sin.}\ }\textbf {\bibinfo {volume} {21}},\ \bibinfo
  {eid} {75} (\bibinfo {year} {1965})}\BibitemShut {NoStop}%
\bibitem [{\citenamefont {Shiba}(1968)}]{Shiba1968}%
  \BibitemOpen
  \bibfield  {author} {\bibinfo {author} {\bibfnamefont {H.}~\bibnamefont
  {Shiba}},\ }\href {\doibase 10.1143/PTP.40.435} {\bibfield  {journal}
  {\bibinfo  {journal} {Prog. Theor. Phys.}\ }\textbf {\bibinfo {volume}
  {40}},\ \bibinfo {pages} {435} (\bibinfo {year} {1968})}\BibitemShut
  {NoStop}%
\bibitem [{\citenamefont {Rusinov}(1968)}]{Rusinov1969}%
  \BibitemOpen
  \bibfield  {author} {\bibinfo {author} {\bibfnamefont {A.~I.}\ \bibnamefont
  {Rusinov}},\ }\href@noop {} {\bibfield  {journal} {\bibinfo  {journal} {Zh.
  Eksp. Teor. Fiz. Pisma Red.}\ }\textbf {\bibinfo {volume} {9}},\ \bibinfo
  {pages} {146} (\bibinfo {year} {1968})},\ \bibinfo {note} {[JETP Lett.
  \textbf{9}, 85 (1969)]}\BibitemShut {NoStop}%
\bibitem [{\citenamefont {Zittartz}\ \emph {et~al.}(1972)\citenamefont
  {Zittartz}, \citenamefont {Bringer},\ and\ \citenamefont
  {M\"{u}ller-Hartmann}}]{Zittartz1972}%
  \BibitemOpen
  \bibfield  {author} {\bibinfo {author} {\bibfnamefont {J.}~\bibnamefont
  {Zittartz}}, \bibinfo {author} {\bibfnamefont {A.}~\bibnamefont {Bringer}}, \
  and\ \bibinfo {author} {\bibfnamefont {E.}~\bibnamefont
  {M\"{u}ller-Hartmann}},\ }\href {\doibase 10.1016/0038-1098(72)90056-7}
  {\bibfield  {journal} {\bibinfo  {journal} {Solid State Commun.}\ }\textbf
  {\bibinfo {volume} {10}},\ \bibinfo {pages} {513} (\bibinfo {year}
  {1972})}\BibitemShut {NoStop}%
\bibitem [{\citenamefont {Tsang}\ and\ \citenamefont
  {Ginsberg}(1980)}]{Tsang1980}%
  \BibitemOpen
  \bibfield  {author} {\bibinfo {author} {\bibfnamefont {J.~K.}\ \bibnamefont
  {Tsang}}\ and\ \bibinfo {author} {\bibfnamefont {D.~M.}\ \bibnamefont
  {Ginsberg}},\ }\href {\doibase 10.1103/PhysRevB.22.4280} {\bibfield
  {journal} {\bibinfo  {journal} {Phys. Rev. B}\ }\textbf {\bibinfo {volume}
  {22}},\ \bibinfo {pages} {4280} (\bibinfo {year} {1980})}\BibitemShut
  {NoStop}%
\bibitem [{\citenamefont {Bauriedl}\ \emph {et~al.}(1981)\citenamefont
  {Bauriedl}, \citenamefont {Ziemann},\ and\ \citenamefont
  {Buckel}}]{Bauriedl1981}%
  \BibitemOpen
  \bibfield  {author} {\bibinfo {author} {\bibfnamefont {W.}~\bibnamefont
  {Bauriedl}}, \bibinfo {author} {\bibfnamefont {P.}~\bibnamefont {Ziemann}}, \
  and\ \bibinfo {author} {\bibfnamefont {W.}~\bibnamefont {Buckel}},\ }\href
  {\doibase 10.1103/PhysRevLett.47.1163} {\bibfield  {journal} {\bibinfo
  {journal} {Phys. Rev. Lett.}\ }\textbf {\bibinfo {volume} {47}},\ \bibinfo
  {pages} {1163} (\bibinfo {year} {1981})}\BibitemShut {NoStop}%
\bibitem [{\citenamefont {Balatsky}\ \emph {et~al.}(2006)\citenamefont
  {Balatsky}, \citenamefont {Vekhter},\ and\ \citenamefont
  {Zhu}}]{Balatsky2006}%
  \BibitemOpen
  \bibfield  {author} {\bibinfo {author} {\bibfnamefont {A.~V.}\ \bibnamefont
  {Balatsky}}, \bibinfo {author} {\bibfnamefont {I.}~\bibnamefont {Vekhter}}, \
  and\ \bibinfo {author} {\bibfnamefont {J.-X.}\ \bibnamefont {Zhu}},\ }\href
  {\doibase 10.1103/RevModPhys.78.373} {\bibfield  {journal} {\bibinfo
  {journal} {Rev. Mod. Phys.}\ }\textbf {\bibinfo {volume} {78}},\ \bibinfo
  {pages} {373} (\bibinfo {year} {2006})}\BibitemShut {NoStop}%
\bibitem [{\citenamefont {Kim}\ \emph {et~al.}(2015)\citenamefont {Kim},
  \citenamefont {Zhang}, \citenamefont {Rossi},\ and\ \citenamefont
  {Lutchyn}}]{Kim2015}%
  \BibitemOpen
  \bibfield  {author} {\bibinfo {author} {\bibfnamefont {Y.}~\bibnamefont
  {Kim}}, \bibinfo {author} {\bibfnamefont {J.}~\bibnamefont {Zhang}}, \bibinfo
  {author} {\bibfnamefont {E.}~\bibnamefont {Rossi}}, \ and\ \bibinfo {author}
  {\bibfnamefont {R.~M.}\ \bibnamefont {Lutchyn}},\ }\href {\doibase
  10.1103/PhysRevLett.114.236804} {\bibfield  {journal} {\bibinfo  {journal}
  {Phys. Rev. Lett.}\ }\textbf {\bibinfo {volume} {114}},\ \bibinfo {pages}
  {236804} (\bibinfo {year} {2015})}\BibitemShut {NoStop}%
\bibitem [{\citenamefont {Sakurai}(1970)}]{Sakurai1970}%
  \BibitemOpen
  \bibfield  {author} {\bibinfo {author} {\bibfnamefont {A.}~\bibnamefont
  {Sakurai}},\ }\href {\doibase 10.1143/PTP.44.1472} {\bibfield  {journal}
  {\bibinfo  {journal} {Progr. Theor. Phys.}\ }\textbf {\bibinfo {volume}
  {44}},\ \bibinfo {pages} {1472} (\bibinfo {year} {1970})}\BibitemShut
  {NoStop}%
\bibitem [{\citenamefont {Salkola}\ \emph {et~al.}(1997)\citenamefont
  {Salkola}, \citenamefont {Balatsky},\ and\ \citenamefont
  {Schrieffer}}]{Salkola1997}%
  \BibitemOpen
  \bibfield  {author} {\bibinfo {author} {\bibfnamefont {M.~I.}\ \bibnamefont
  {Salkola}}, \bibinfo {author} {\bibfnamefont {A.~V.}\ \bibnamefont
  {Balatsky}}, \ and\ \bibinfo {author} {\bibfnamefont {J.~R.}\ \bibnamefont
  {Schrieffer}},\ }\href {\doibase 10.1103/PhysRevB.55.12648} {\bibfield
  {journal} {\bibinfo  {journal} {Phys. Rev. B}\ }\textbf {\bibinfo {volume}
  {55}},\ \bibinfo {pages} {12648} (\bibinfo {year} {1997})}\BibitemShut
  {NoStop}%
\bibitem [{\citenamefont {Morr}\ and\ \citenamefont
  {Stavropoulos}(2003)}]{Morr2003}%
  \BibitemOpen
  \bibfield  {author} {\bibinfo {author} {\bibfnamefont {D.~K.}\ \bibnamefont
  {Morr}}\ and\ \bibinfo {author} {\bibfnamefont {N.~A.}\ \bibnamefont
  {Stavropoulos}},\ }\href {\doibase 10.1103/PhysRevB.67.020502} {\bibfield
  {journal} {\bibinfo  {journal} {Phys. Rev. B}\ }\textbf {\bibinfo {volume}
  {67}},\ \bibinfo {pages} {020502} (\bibinfo {year} {2003})}\BibitemShut
  {NoStop}%
\bibitem [{\citenamefont {Morr}\ and\ \citenamefont {Yoon}(2006)}]{Morr2006}%
  \BibitemOpen
  \bibfield  {author} {\bibinfo {author} {\bibfnamefont {D.~K.}\ \bibnamefont
  {Morr}}\ and\ \bibinfo {author} {\bibfnamefont {J.}~\bibnamefont {Yoon}},\
  }\href {\doibase 10.1103/PhysRevB.73.224511} {\bibfield  {journal} {\bibinfo
  {journal} {Phys. Rev. B}\ }\textbf {\bibinfo {volume} {73}},\ \bibinfo
  {pages} {224511} (\bibinfo {year} {2006})}\BibitemShut {NoStop}%
\bibitem [{\citenamefont {Kondo}(1964)}]{Kondo1964}%
  \BibitemOpen
  \bibfield  {author} {\bibinfo {author} {\bibfnamefont {J.}~\bibnamefont
  {Kondo}},\ }\href {\doibase 10.1143/PTP.32.37} {\bibfield  {journal}
  {\bibinfo  {journal} {Progr. Theor. Phys.}\ }\textbf {\bibinfo {volume}
  {32}},\ \bibinfo {pages} {37} (\bibinfo {year} {1964})}\BibitemShut {NoStop}%
\bibitem [{\citenamefont {Liu}(1965)}]{Liu1965}%
  \BibitemOpen
  \bibfield  {author} {\bibinfo {author} {\bibfnamefont {S.~H.}\ \bibnamefont
  {Liu}},\ }\href {\doibase 10.1103/PhysRev.137.A1209} {\bibfield  {journal}
  {\bibinfo  {journal} {Phys. Rev.}\ }\textbf {\bibinfo {volume} {137}},\
  \bibinfo {pages} {A1209} (\bibinfo {year} {1965})}\BibitemShut {NoStop}%
\bibitem [{\citenamefont {Griffin}(1965)}]{Griffin1965}%
  \BibitemOpen
  \bibfield  {author} {\bibinfo {author} {\bibfnamefont {A.}~\bibnamefont
  {Griffin}},\ }\href {\doibase 10.1103/PhysRevLett.15.703} {\bibfield
  {journal} {\bibinfo  {journal} {Phys. Rev. Lett.}\ }\textbf {\bibinfo
  {volume} {15}},\ \bibinfo {pages} {703} (\bibinfo {year} {1965})}\BibitemShut
  {NoStop}%
\bibitem [{\citenamefont {Maki}(1967)}]{Maki1967}%
  \BibitemOpen
  \bibfield  {author} {\bibinfo {author} {\bibfnamefont {K.}~\bibnamefont
  {Maki}},\ }\href {\doibase 10.1103/PhysRev.153.428} {\bibfield  {journal}
  {\bibinfo  {journal} {Phys. Rev.}\ }\textbf {\bibinfo {volume} {153}},\
  \bibinfo {pages} {428} (\bibinfo {year} {1967})}\BibitemShut {NoStop}%
\bibitem [{\citenamefont {Soda}\ \emph {et~al.}(1967)\citenamefont {Soda},
  \citenamefont {Matsuura},\ and\ \citenamefont {Nagaoka}}]{Soda1967}%
  \BibitemOpen
  \bibfield  {author} {\bibinfo {author} {\bibfnamefont {T.}~\bibnamefont
  {Soda}}, \bibinfo {author} {\bibfnamefont {T.}~\bibnamefont {Matsuura}}, \
  and\ \bibinfo {author} {\bibfnamefont {Y.}~\bibnamefont {Nagaoka}},\ }\href
  {\doibase 10.1143/PTP.38.551} {\bibfield  {journal} {\bibinfo  {journal}
  {Progr. Theor. Phys.}\ }\textbf {\bibinfo {volume} {38}},\ \bibinfo {pages}
  {551} (\bibinfo {year} {1967})}\BibitemShut {NoStop}%
\bibitem [{\citenamefont {Fowler}\ and\ \citenamefont
  {Maki}(1967)}]{Fowler1967}%
  \BibitemOpen
  \bibfield  {author} {\bibinfo {author} {\bibfnamefont {M.}~\bibnamefont
  {Fowler}}\ and\ \bibinfo {author} {\bibfnamefont {K.}~\bibnamefont {Maki}},\
  }\href {\doibase 10.1103/PhysRev.164.484} {\bibfield  {journal} {\bibinfo
  {journal} {Phys. Rev.}\ }\textbf {\bibinfo {volume} {164}},\ \bibinfo {pages}
  {484} (\bibinfo {year} {1967})}\BibitemShut {NoStop}%
\bibitem [{\citenamefont {Abrikosov}(1969)}]{Abrikosov1969}%
  \BibitemOpen
  \bibfield  {author} {\bibinfo {author} {\bibfnamefont {A.~A.}\ \bibnamefont
  {Abrikosov}},\ }\href {\doibase 10.1070/PU1969v012n02ABEH003930} {\bibfield
  {journal} {\bibinfo  {journal} {Sov. Phys. Usp.}\ }\textbf {\bibinfo {volume}
  {12}},\ \bibinfo {pages} {168} (\bibinfo {year} {1969})}\BibitemShut
  {NoStop}%
\bibitem [{\citenamefont {Takano}\ and\ \citenamefont
  {Matayoshi}(1969)}]{Takano1969}%
  \BibitemOpen
  \bibfield  {author} {\bibinfo {author} {\bibfnamefont {F.}~\bibnamefont
  {Takano}}\ and\ \bibinfo {author} {\bibfnamefont {S.}~\bibnamefont
  {Matayoshi}},\ }\href {\doibase 10.1143/PTP.41.45} {\bibfield  {journal}
  {\bibinfo  {journal} {Progr. Theor. Phys.}\ }\textbf {\bibinfo {volume}
  {41}},\ \bibinfo {pages} {45} (\bibinfo {year} {1969})}\BibitemShut {NoStop}%
\bibitem [{\citenamefont {Kitamura}(1970)}]{Kitamura1970}%
  \BibitemOpen
  \bibfield  {author} {\bibinfo {author} {\bibfnamefont {T.}~\bibnamefont
  {Kitamura}},\ }\href {\doibase 10.1143/PTP.43.271} {\bibfield  {journal}
  {\bibinfo  {journal} {Progr. Theor. Phys.}\ }\textbf {\bibinfo {volume}
  {43}},\ \bibinfo {pages} {271} (\bibinfo {year} {1970})}\BibitemShut
  {NoStop}%
\bibitem [{\citenamefont {Fowler}\ and\ \citenamefont
  {Maki}(1970)}]{Fowler1970}%
  \BibitemOpen
  \bibfield  {author} {\bibinfo {author} {\bibfnamefont {M.}~\bibnamefont
  {Fowler}}\ and\ \bibinfo {author} {\bibfnamefont {K.}~\bibnamefont {Maki}},\
  }\href {\doibase 10.1103/PhysRevB.1.181} {\bibfield  {journal} {\bibinfo
  {journal} {Phys. Rev. B}\ }\textbf {\bibinfo {volume} {1}},\ \bibinfo {pages}
  {181} (\bibinfo {year} {1970})}\BibitemShut {NoStop}%
\bibitem [{\citenamefont {M\"{u}ller-Hartmann}\ and\ \citenamefont
  {Zittartz}(1971)}]{MullerHartmann1971}%
  \BibitemOpen
  \bibfield  {author} {\bibinfo {author} {\bibfnamefont {E.}~\bibnamefont
  {M\"{u}ller-Hartmann}}\ and\ \bibinfo {author} {\bibfnamefont
  {J.}~\bibnamefont {Zittartz}},\ }\href {\doibase 10.1103/PhysRevLett.26.428}
  {\bibfield  {journal} {\bibinfo  {journal} {Phys. Rev. Lett.}\ }\textbf
  {\bibinfo {volume} {26}},\ \bibinfo {pages} {428} (\bibinfo {year}
  {1971})}\BibitemShut {NoStop}%
\bibitem [{\citenamefont {Matsuura}\ \emph {et~al.}(1977)\citenamefont
  {Matsuura}, \citenamefont {Ichinose},\ and\ \citenamefont
  {Nagaoka}}]{Matsuura1977}%
  \BibitemOpen
  \bibfield  {author} {\bibinfo {author} {\bibfnamefont {T.}~\bibnamefont
  {Matsuura}}, \bibinfo {author} {\bibfnamefont {S.}~\bibnamefont {Ichinose}},
  \ and\ \bibinfo {author} {\bibfnamefont {Y.}~\bibnamefont {Nagaoka}},\ }\href
  {\doibase 10.1143/PTP.57.713} {\bibfield  {journal} {\bibinfo  {journal}
  {Progr. Theor. Phys.}\ }\textbf {\bibinfo {volume} {57}},\ \bibinfo {pages}
  {713} (\bibinfo {year} {1977})}\BibitemShut {NoStop}%
\bibitem [{\citenamefont {Ichinose}(1977{\natexlab{a}})}]{Ichinose1977}%
  \BibitemOpen
  \bibfield  {author} {\bibinfo {author} {\bibfnamefont {S.}~\bibnamefont
  {Ichinose}},\ }\href {\doibase 10.1143/PTP.58.404} {\bibfield  {journal}
  {\bibinfo  {journal} {Progr. Theor. Phys.}\ }\textbf {\bibinfo {volume}
  {58}},\ \bibinfo {pages} {404} (\bibinfo {year}
  {1977}{\natexlab{a}})}\BibitemShut {NoStop}%
\bibitem [{\citenamefont {Ichinose}(1977{\natexlab{b}})}]{Ichinose1977_2}%
  \BibitemOpen
  \bibfield  {author} {\bibinfo {author} {\bibfnamefont {S.}~\bibnamefont
  {Ichinose}},\ }\href {\doibase 10.1143/PTP.58.733} {\bibfield  {journal}
  {\bibinfo  {journal} {Progr. Theor. Phys.}\ }\textbf {\bibinfo {volume}
  {58}},\ \bibinfo {pages} {733} (\bibinfo {year}
  {1977}{\natexlab{b}})}\BibitemShut {NoStop}%
\bibitem [{\citenamefont {Franke}\ \emph {et~al.}(2011)\citenamefont {Franke},
  \citenamefont {Schulze},\ and\ \citenamefont {Pascual}}]{Franke2011}%
  \BibitemOpen
  \bibfield  {author} {\bibinfo {author} {\bibfnamefont {K.~J.}\ \bibnamefont
  {Franke}}, \bibinfo {author} {\bibfnamefont {G.}~\bibnamefont {Schulze}}, \
  and\ \bibinfo {author} {\bibfnamefont {J.~I.}\ \bibnamefont {Pascual}},\
  }\href {\doibase 10.1126/science.1202204} {\bibfield  {journal} {\bibinfo
  {journal} {Science}\ }\textbf {\bibinfo {volume} {332}},\ \bibinfo {pages}
  {940} (\bibinfo {year} {2011})}\BibitemShut {NoStop}%
\bibitem [{\citenamefont {Gor'kov}\ and\ \citenamefont
  {A.~I.~Rusinov}(1964)}]{Gorkov_Rusinov_1964}%
  \BibitemOpen
  \bibfield  {author} {\bibinfo {author} {\bibfnamefont {L.~P.}\ \bibnamefont
  {Gor'kov}}\ and\ \bibinfo {author} {\bibfnamefont {A.~I.}\ \bibnamefont
  {A.~I.~Rusinov}},\ }\href@noop {} {\bibfield  {journal} {\bibinfo  {journal}
  {Zh. Eksp. Teor. Fiz.}\ }\textbf {\bibinfo {volume} {46}},\ \bibinfo {pages}
  {1363} (\bibinfo {year} {1964})},\ \bibinfo {note} {[Sov. Phys.--JETP {\bf
  19}, 922 (1964)]}\BibitemShut {NoStop}%
\bibitem [{\citenamefont {Fulde}\ and\ \citenamefont
  {Maki}(1966)}]{Fulde_Maki_1966}%
  \BibitemOpen
  \bibfield  {author} {\bibinfo {author} {\bibfnamefont {P.}~\bibnamefont
  {Fulde}}\ and\ \bibinfo {author} {\bibfnamefont {K.}~\bibnamefont {Maki}},\
  }\href {\doibase 10.1103/PhysRev.141.275} {\bibfield  {journal} {\bibinfo
  {journal} {Phys. Rev.}\ }\textbf {\bibinfo {volume} {141}},\ \bibinfo {pages}
  {275} (\bibinfo {year} {1966})}\BibitemShut {NoStop}%
\bibitem [{\citenamefont {Izyumov}\ and\ \citenamefont
  {Skryabin}(1974)}]{Izyumov1974}%
  \BibitemOpen
  \bibfield  {author} {\bibinfo {author} {\bibfnamefont {Y.~A.}\ \bibnamefont
  {Izyumov}}\ and\ \bibinfo {author} {\bibfnamefont {Y.~N.}\ \bibnamefont
  {Skryabin}},\ }\href {\doibase 10.1002/pssb.2220610102} {\bibfield  {journal}
  {\bibinfo  {journal} {Phys. Status Solidi B}\ }\textbf {\bibinfo {volume}
  {61}},\ \bibinfo {pages} {9} (\bibinfo {year} {1974})}\BibitemShut {NoStop}%
\bibitem [{\citenamefont {Rammer}(1998)}]{Rammer_book1998}%
  \BibitemOpen
  \bibfield  {author} {\bibinfo {author} {\bibfnamefont {J.}~\bibnamefont
  {Rammer}},\ }\href@noop {} {\emph {\bibinfo {title} {{Quantum Transport
  Theory}}}}\ (\bibinfo  {publisher} {Perseus Books, Reading},\ \bibinfo
  {address} {MA},\ \bibinfo {year} {1998})\BibitemShut {NoStop}%
\bibitem [{\citenamefont {Eilenberger}(1968)}]{Eilenberger1968}%
  \BibitemOpen
  \bibfield  {author} {\bibinfo {author} {\bibfnamefont {G.}~\bibnamefont
  {Eilenberger}},\ }\href@noop {} {\bibfield  {journal} {\bibinfo  {journal}
  {Z. Phys.}\ }\textbf {\bibinfo {volume} {214}},\ \bibinfo {pages} {195}
  (\bibinfo {year} {1968})}\BibitemShut {NoStop}%
\bibitem [{\citenamefont {Larkin}\ and\ \citenamefont
  {Ovchinnikov}(1968)}]{Larkin_Ovchinnikov_1969}%
  \BibitemOpen
  \bibfield  {author} {\bibinfo {author} {\bibfnamefont {A.~I.}\ \bibnamefont
  {Larkin}}\ and\ \bibinfo {author} {\bibfnamefont {{\relax Yu}.~N.}\
  \bibnamefont {Ovchinnikov}},\ }\href@noop {} {\bibfield  {journal} {\bibinfo
  {journal} {Zh. Eksp. Teor. Fiz.}\ }\textbf {\bibinfo {volume} {55}},\
  \bibinfo {pages} {2262} (\bibinfo {year} {1968})},\ \bibinfo {note} {[Sov.
  Phys.--JETP {\bf 28}, 1200 (1969)]}\BibitemShut {NoStop}%
\bibitem [{\citenamefont {Eschrig}(2000)}]{Eschrig_PRB2000}%
  \BibitemOpen
  \bibfield  {author} {\bibinfo {author} {\bibfnamefont {M.}~\bibnamefont
  {Eschrig}},\ }\href {\doibase 10.1103/PhysRevB.61.9061} {\bibfield  {journal}
  {\bibinfo  {journal} {Phys. Rev. B}\ }\textbf {\bibinfo {volume} {61}},\
  \bibinfo {pages} {9061} (\bibinfo {year} {2000})}\BibitemShut {NoStop}%
\bibitem [{\citenamefont {Eschrig}(2009)}]{Eschrig_PRB2009}%
  \BibitemOpen
  \bibfield  {author} {\bibinfo {author} {\bibfnamefont {M.}~\bibnamefont
  {Eschrig}},\ }\href {\doibase 10.1103/PhysRevB.80.134511} {\bibfield
  {journal} {\bibinfo  {journal} {Phys. Rev. B}\ }\textbf {\bibinfo {volume}
  {80}},\ \bibinfo {pages} {134511} (\bibinfo {year} {2009})}\BibitemShut
  {NoStop}%
\bibitem [{Note1()}]{Note1}%
  \BibitemOpen
  \bibinfo {note} {Because of the point-like scattering potential of impurities
  $v_{\protect \mathrm {S}} = 2\protect \mathaccentV {tilde}07E{v}_{\protect
  \mathrm {S}}/3$, where $\protect \mathaccentV {tilde}07E{v}_{\protect \mathrm
  {S}} = g\mu _B\mu _0\protect \mathcal {M}/2$. Here $g$ is the quasiparticle
  g-factor, $\mu _B$ is the Bohr magneton, $\mu _0$ is the vacuum permeability,
  and $\protect \mathcal {M}$ is the magnitude of the impurity magnetic moment.
  The factor $2/3$ comes from taking into account the $\protect \mathbf
  {H}$-field of a point-like magnetic dipole.}\BibitemShut {Stop}%
\bibitem [{Note2()}]{Note2}%
  \BibitemOpen
  \bibinfo {note} {It might seem counterintuitive that anti-ferromagnetic
  interaction is described by $\alpha > 0$, however it is easily understood.
  Magnetic moment of an electron is $\protect \bm {\mu }_e = -g\mu _B\protect
  \bm {\sigma }/2$, while its spin angular momentum is $\protect \mathbf {s}_e
  = \hbar \protect \bm {\sigma }/2$. Since for $\alpha >0$ the itinerant
  electrons interact anti-ferromagnetically with the impurity spin $\protect
  \mathbf {S}\propto \protect \mathbf {s}_e$, see Fig. \ref {Fig2}, the
  interaction is ferromagnetic in terms of the impurity magnetic moment
  $\protect \bm {\protect \mathcal {M}}\propto \protect \bm {\mu
  }_e$.}\BibitemShut {Stop}%
\bibitem [{\citenamefont {Matsubara}(1955)}]{Matsubara1955}%
  \BibitemOpen
  \bibfield  {author} {\bibinfo {author} {\bibfnamefont {T.}~\bibnamefont
  {Matsubara}},\ }\href {\doibase 10.1143/PTP.14.351} {\bibfield  {journal}
  {\bibinfo  {journal} {Progr. Theor. Phys.}\ }\textbf {\bibinfo {volume}
  {14}},\ \bibinfo {pages} {351} (\bibinfo {year} {1955})}\BibitemShut
  {NoStop}%
\bibitem [{\citenamefont {Schopohl}\ and\ \citenamefont
  {Maki}(1995)}]{Schopohl_Maki_PRB1995}%
  \BibitemOpen
  \bibfield  {author} {\bibinfo {author} {\bibfnamefont {N.}~\bibnamefont
  {Schopohl}}\ and\ \bibinfo {author} {\bibfnamefont {K.}~\bibnamefont
  {Maki}},\ }\href {\doibase 10.1103/PhysRevB.52.490} {\bibfield  {journal}
  {\bibinfo  {journal} {Phys. Rev. B}\ }\textbf {\bibinfo {volume} {52}},\
  \bibinfo {pages} {490} (\bibinfo {year} {1995})}\BibitemShut {NoStop}%
\bibitem [{\citenamefont {Schopohl}()}]{Schopohl_1998}%
  \BibitemOpen
  \bibfield  {author} {\bibinfo {author} {\bibfnamefont {N.}~\bibnamefont
  {Schopohl}},\ }\href@noop {} {\enquote {\bibinfo {title} {{Transformation of
  the Eilenberger Equations of Superconductivity to a Scalar Riccati
  Equation}},}\ }\Eprint {http://arxiv.org/abs/arXiv:cond-mat/9804064}
  {arXiv:cond-mat/9804064} \BibitemShut {NoStop}%
\bibitem [{\citenamefont {Grein}\ \emph {et~al.}(2013)\citenamefont {Grein},
  \citenamefont {L\"{o}fwander},\ and\ \citenamefont
  {Eschrig}}]{Grein_PRB2013}%
  \BibitemOpen
  \bibfield  {author} {\bibinfo {author} {\bibfnamefont {R.}~\bibnamefont
  {Grein}}, \bibinfo {author} {\bibfnamefont {T.}~\bibnamefont
  {L\"{o}fwander}}, \ and\ \bibinfo {author} {\bibfnamefont {M.}~\bibnamefont
  {Eschrig}},\ }\href {\doibase 10.1103/PhysRevB.88.054502} {\bibfield
  {journal} {\bibinfo  {journal} {Phys. Rev. B}\ }\textbf {\bibinfo {volume}
  {88}},\ \bibinfo {pages} {054502} (\bibinfo {year} {2013})}\BibitemShut
  {NoStop}%
\bibitem [{\citenamefont {Baym}\ and\ \citenamefont {Mermin}(1961)}]{Baym1961}%
  \BibitemOpen
  \bibfield  {author} {\bibinfo {author} {\bibfnamefont {G.}~\bibnamefont
  {Baym}}\ and\ \bibinfo {author} {\bibfnamefont {N.~D.}\ \bibnamefont
  {Mermin}},\ }\href {\doibase 10.1063/1.1703704} {\bibfield  {journal}
  {\bibinfo  {journal} {J. Math. Phys.}\ }\textbf {\bibinfo {volume} {2}},\
  \bibinfo {pages} {232} (\bibinfo {year} {1961})}\BibitemShut {NoStop}%
\bibitem [{\citenamefont {Serene}\ and\ \citenamefont
  {Rainer}(1983)}]{Serene_Rainer_PhysRep1983}%
  \BibitemOpen
  \bibfield  {author} {\bibinfo {author} {\bibfnamefont {J.~W.}\ \bibnamefont
  {Serene}}\ and\ \bibinfo {author} {\bibfnamefont {D.}~\bibnamefont
  {Rainer}},\ }\href {\doibase 10.1016/0370-1573(83)90051-0"} {\bibfield
  {journal} {\bibinfo  {journal} {Phys. Rep.}\ }\textbf {\bibinfo {volume}
  {101}},\ \bibinfo {pages} {221} (\bibinfo {year} {1983})}\BibitemShut
  {NoStop}%
\bibitem [{Note3()}]{Note3}%
  \BibitemOpen
  \bibinfo {note} {In order for these results to be applicable to a given
  sample with a given distribution of impurities, the size of the system must
  be much larger than the phase coherence length. Then, the system is said to
  be self-averaging. For the case of small superconducting islands one has to
  consider the actual spatial arrangement of impurities to make sensible
  predictions.}\BibitemShut {Stop}%
\bibitem [{\citenamefont {Gor'kov}(2008)}]{Bennemann_Ketterson_2008}%
  \BibitemOpen
  \bibfield  {author} {\bibinfo {author} {\bibfnamefont {L.~P.}\ \bibnamefont
  {Gor'kov}},\ }in\ \href@noop {} {\emph {\bibinfo {booktitle}
  {{Superconductivity: Conventional and Unconventional Superconductors}}}},\
  \bibinfo {editor} {edited by\ \bibinfo {editor} {\bibfnamefont {K.~H.}\
  \bibnamefont {Bennemann}}\ and\ \bibinfo {editor} {\bibfnamefont {J.~B.}\
  \bibnamefont {Ketterson}}}\ (\bibinfo  {publisher} {Springer-Verlag},\
  \bibinfo {address} {Berlin},\ \bibinfo {year} {2008})\BibitemShut {NoStop}%
\bibitem [{Note4()}]{Note4}%
  \BibitemOpen
  \bibinfo {note} {A crucial feature of the quasiclassical theory is that there
  exists a separation of energy (or length) scales, e.g. $\Delta _0\ll E_F$,
  where $E_F$ is the Fermi energy. This enables one to introduce a small
  parameter $\protect \mathtt {small}$, which is used as an expansion parameter
  for the full microscopic propagators and self-energies \cite
  {Serene_Rainer_PhysRep1983,Sauls_FLT1994}. In our case, apart from the order
  parameter being small compared to $E_F$ we have to assume that
  $nv_{0,\protect \mathrm {S}}\ll E_F$. Following the procedure of calculating
  physical observables described in Ref. [\protect \rev@citealpnum
  {Serene_Rainer_PhysRep1983}] we have obtained Eq. (\ref {M_eqn}). The first
  term is the so-called high-energy correction, which is not captured by
  quasiclassics and has to be computed separately.}\BibitemShut {Stop}%
\bibitem [{\citenamefont {Keller}\ \emph {et~al.}(1988)\citenamefont {Keller},
  \citenamefont {Scharnberg},\ and\ \citenamefont {Monien}}]{Keller1988}%
  \BibitemOpen
  \bibfield  {author} {\bibinfo {author} {\bibfnamefont {J.}~\bibnamefont
  {Keller}}, \bibinfo {author} {\bibfnamefont {K.}~\bibnamefont {Scharnberg}},
  \ and\ \bibinfo {author} {\bibfnamefont {H.}~\bibnamefont {Monien}},\ }\href
  {\doibase 10.1016/0921-4534(88)90088-3} {\bibfield  {journal} {\bibinfo
  {journal} {Physica C}\ }\textbf {\bibinfo {volume} {152}},\ \bibinfo {pages}
  {302} (\bibinfo {year} {1988})}\BibitemShut {NoStop}%
\bibitem [{\citenamefont {Sauls}(1994)}]{Sauls_FLT1994}%
  \BibitemOpen
  \bibfield  {author} {\bibinfo {author} {\bibfnamefont {J.~A.}\ \bibnamefont
  {Sauls}},\ }in\ \href@noop {} {\emph {\bibinfo {booktitle} {{Strongly
  Correlated Electronic Materials: The Los Alamos Symposium 1993}}}},\ \bibinfo
  {editor} {edited by\ \bibinfo {editor} {\bibfnamefont {K.~S.}\ \bibnamefont
  {Kevin S.~Bedell}}, \bibinfo {editor} {\bibfnamefont {Z.}~\bibnamefont
  {Wang}}, \bibinfo {editor} {\bibfnamefont {D.~E.}\ \bibnamefont {Meltzer}},
  \bibinfo {editor} {\bibfnamefont {A.~V.}\ \bibnamefont {Balatsky}}, \ and\
  \bibinfo {editor} {\bibfnamefont {E.}~\bibnamefont {Abrahams}}}\ (\bibinfo
  {publisher} {Addison-Wesely, Reading},\ \bibinfo {address} {MA},\ \bibinfo
  {year} {1994})\BibitemShut {NoStop}%
\bibitem [{\citenamefont {Okabe}\ and\ \citenamefont {Nagi}(1983)}]{Okabe1983}%
  \BibitemOpen
  \bibfield  {author} {\bibinfo {author} {\bibfnamefont {Y.}~\bibnamefont
  {Okabe}}\ and\ \bibinfo {author} {\bibfnamefont {A.~D.~S.}\ \bibnamefont
  {Nagi}},\ }\href {\doibase 10.1103/PhysRevB.28.1320} {\bibfield  {journal}
  {\bibinfo  {journal} {Phys. Rev. B}\ }\textbf {\bibinfo {volume} {28}},\
  \bibinfo {pages} {1320} (\bibinfo {year} {1983})}\BibitemShut {NoStop}%
\bibitem [{\citenamefont {Clogston}(1962)}]{Clogston1962}%
  \BibitemOpen
  \bibfield  {author} {\bibinfo {author} {\bibfnamefont {A.~M.}\ \bibnamefont
  {Clogston}},\ }\href {\doibase 10.1103/PhysRevLett.9.266} {\bibfield
  {journal} {\bibinfo  {journal} {Phys. Rev. Lett.}\ }\textbf {\bibinfo
  {volume} {9}},\ \bibinfo {pages} {266} (\bibinfo {year} {1962})}\BibitemShut
  {NoStop}%
\bibitem [{\citenamefont {Chandrasekhar}(1962)}]{Chandrasekhar1962}%
  \BibitemOpen
  \bibfield  {author} {\bibinfo {author} {\bibfnamefont {B.~S.}\ \bibnamefont
  {Chandrasekhar}},\ }\href {\doibase 10.1063/1.1777362} {\bibfield  {journal}
  {\bibinfo  {journal} {Appl. Phys. Lett.}\ }\textbf {\bibinfo {volume} {1}},\
  \bibinfo {pages} {7} (\bibinfo {year} {1962})}\BibitemShut {NoStop}%
\bibitem [{\citenamefont {Fulde}\ and\ \citenamefont
  {Ferrell}(1964)}]{Fulde_Ferrel_1964}%
  \BibitemOpen
  \bibfield  {author} {\bibinfo {author} {\bibfnamefont {P.}~\bibnamefont
  {Fulde}}\ and\ \bibinfo {author} {\bibfnamefont {R.~A.}\ \bibnamefont
  {Ferrell}},\ }\href {\doibase 10.1103/PhysRev.135.A550} {\bibfield  {journal}
  {\bibinfo  {journal} {Phys. Rev.}\ }\textbf {\bibinfo {volume} {135}},\
  \bibinfo {pages} {A550} (\bibinfo {year} {1964})}\BibitemShut {NoStop}%
\bibitem [{\citenamefont {Larkin}\ and\ \citenamefont
  {Ovchinnikov}(1964)}]{Larkin_Ovchinnikov_1964}%
  \BibitemOpen
  \bibfield  {author} {\bibinfo {author} {\bibfnamefont {A.~I.}\ \bibnamefont
  {Larkin}}\ and\ \bibinfo {author} {\bibfnamefont {{\relax Yu}.~N.}\
  \bibnamefont {Ovchinnikov}},\ }\href@noop {} {\bibfield  {journal} {\bibinfo
  {journal} {Zh. Eksp. Teor. Fiz.}\ }\textbf {\bibinfo {volume} {47}},\
  \bibinfo {pages} {1136} (\bibinfo {year} {1964})},\ \bibinfo {note} {[Sov.
  Phys.--JETP {\bf 20}, 762 (1965)]}\BibitemShut {NoStop}%
\bibitem [{Note5()}]{Note5}%
  \BibitemOpen
  \bibinfo {note} {There is a small range of parameters where a gapless
  superconductivity \cite {Abrikosov_Gorkov_1961} is possible. Nevertheless,
  roughly speaking, transition to the normal state occurs when the impurity
  band fills in the energy gap.}\BibitemShut {Stop}%
\bibitem [{\citenamefont {Shevtsov}\ and\ \citenamefont
  {T~L\"{o}fwander}(2014)}]{Shevtsov_LT2014}%
  \BibitemOpen
  \bibfield  {author} {\bibinfo {author} {\bibfnamefont {O.}~\bibnamefont
  {Shevtsov}}\ and\ \bibinfo {author} {\bibfnamefont {T.}~\bibnamefont
  {T~L\"{o}fwander}},\ }\href {\doibase 10.1088/1742-6596/568/2/022044}
  {\bibfield  {journal} {\bibinfo  {journal} {J. Phys.: Conf. Ser.}\ }\textbf
  {\bibinfo {volume} {568}},\ \bibinfo {pages} {022044} (\bibinfo {year}
  {2014})}\BibitemShut {NoStop}%
\bibitem [{\citenamefont {Abrikosov}\ and\ \citenamefont
  {Gor'kov}(1963)}]{Abrikosov_Gorkov_1963}%
  \BibitemOpen
  \bibfield  {author} {\bibinfo {author} {\bibfnamefont {A.~A.}\ \bibnamefont
  {Abrikosov}}\ and\ \bibinfo {author} {\bibfnamefont {L.~P.}\ \bibnamefont
  {Gor'kov}},\ }\href@noop {} {\bibfield  {journal} {\bibinfo  {journal} {Zh.
  Eksp. Teor. Fiz.}\ }\textbf {\bibinfo {volume} {43}},\ \bibinfo {pages}
  {2230} (\bibinfo {year} {1963})},\ \bibinfo {note} {[Sov. Phys.--JETP {\bf
  16}, 1575 (1963)]}\BibitemShut {NoStop}%
\bibitem [{\citenamefont {Anderson}(1959)}]{Anderson1956}%
  \BibitemOpen
  \bibfield  {author} {\bibinfo {author} {\bibfnamefont {P.~W.}\ \bibnamefont
  {Anderson}},\ }\href@noop {} {\bibfield  {journal} {\bibinfo  {journal} {J.
  Phys. Chem. Solids}\ }\textbf {\bibinfo {volume} {11}},\ \bibinfo {pages}
  {26} (\bibinfo {year} {1959})}\BibitemShut {NoStop}%
\bibitem [{Note6()}]{Note6}%
  \BibitemOpen
  \bibinfo {note} {Note that the same result also holds for unpolarized
  magnetic impurities. We think that even if there is no net magnetic field in
  this case, the local exchange field created by each individual impurity would
  destroy superconductivity before the unitary limit is reached.}\BibitemShut
  {Stop}%
\bibitem [{\citenamefont {Balatsky}\ and\ \citenamefont
  {Trugman}(1997)}]{Balatsky1997}%
  \BibitemOpen
  \bibfield  {author} {\bibinfo {author} {\bibfnamefont {A.~V.}\ \bibnamefont
  {Balatsky}}\ and\ \bibinfo {author} {\bibfnamefont {S.~A.}\ \bibnamefont
  {Trugman}},\ }\href {\doibase 10.1103/PhysRevLett.79.3767} {\bibfield
  {journal} {\bibinfo  {journal} {Phys. Rev. Lett.}\ }\textbf {\bibinfo
  {volume} {79}},\ \bibinfo {pages} {3767} (\bibinfo {year}
  {1997})}\BibitemShut {NoStop}%
\bibitem [{\citenamefont {Kim}\ \emph {et~al.}(2014)\citenamefont {Kim},
  \citenamefont {Lee},\ and\ \citenamefont {Hong}}]{Kim2014}%
  \BibitemOpen
  \bibfield  {author} {\bibinfo {author} {\bibfnamefont {S.}~\bibnamefont
  {Kim}}, \bibinfo {author} {\bibfnamefont {S.}~\bibnamefont {Lee}}, \ and\
  \bibinfo {author} {\bibfnamefont {J.}~\bibnamefont {Hong}},\ }\href {\doibase
  10.1021/nn500683b} {\bibfield  {journal} {\bibinfo  {journal} {ACS Nano}\
  }\textbf {\bibinfo {volume} {8}},\ \bibinfo {pages} {4698} (\bibinfo {year}
  {2014})}\BibitemShut {NoStop}%
\bibitem [{Note7()}]{Note7}%
  \BibitemOpen
  \bibinfo {note} {Surprisingly, even when the magnetic impurity is just a
  single atom, the YSR model can still be successfully used to analyze
  experimental data, see Ref. [\protect \rev@citealpnum {Ruby_2015}], even
  though a more complicated physics related to Kondo effect can emerge \cite
  {Liu1965,Griffin1965,Maki1967,Soda1967,Fowler1967,Abrikosov1969,Takano1969,Kitamura1970,Fowler1970,MullerHartmann1971,Matsuura1977,Ichinose1977,Ichinose1977_2,Franke2011}.}\BibitemShut
  {Stop}%
\bibitem [{\citenamefont {Ruby}\ \emph {et~al.}(2015)\citenamefont {Ruby},
  \citenamefont {Pientka}, \citenamefont {Peng}, \citenamefont {von Oppen},
  \citenamefont {Heinrich},\ and\ \citenamefont {Franke}}]{Ruby_2015}%
  \BibitemOpen
  \bibfield  {author} {\bibinfo {author} {\bibfnamefont {M.}~\bibnamefont
  {Ruby}}, \bibinfo {author} {\bibfnamefont {F.}~\bibnamefont {Pientka}},
  \bibinfo {author} {\bibfnamefont {Y.}~\bibnamefont {Peng}}, \bibinfo {author}
  {\bibfnamefont {F.}~\bibnamefont {von Oppen}}, \bibinfo {author}
  {\bibfnamefont {B.~W.}\ \bibnamefont {Heinrich}}, \ and\ \bibinfo {author}
  {\bibfnamefont {K.~J.}\ \bibnamefont {Franke}},\ }\href {\doibase
  10.1103/PhysRevLett.115.087001} {\bibfield  {journal} {\bibinfo  {journal}
  {Phys. Rev. Lett.}\ }\textbf {\bibinfo {volume} {115}},\ \bibinfo {pages}
  {087001} (\bibinfo {year} {2015})}\BibitemShut {NoStop}%
\end{thebibliography}%
\end{document}